\documentclass[a4paper,11pt]{article}

\usepackage[a4paper,
  left=2.5cm, right=2.5cm,
  top= 3cm, bottom=4cm]{geometry}

\usepackage{amsmath,amssymb,amsfonts}
\usepackage{hyperref}
\usepackage{bbm}
\usepackage{epic}
\usepackage{longtable}

\newcommand{\be}{\begin{equation}}  
\newcommand{\ee}{\end{equation}}

\newcommand{\nn}{\nonumber}
\newcommand{\rem}[1]{} 

\def\C{\mathbb{C}}
\def\Z{\mathbb{Z}}
\def\Q{\mathbb{Q}}
\def\R{\mathbb{R}}
\def\N{\mathbb{N}}
\def\P{\mathbb{P}}

\def\Hirz[#1]{\mathbbm{F}_{#1}}
\def\o[#1]{\overline{#1}}

\makeatletter
\newcommand\xleftrightarrow[2][]{%
  \ext@arrow 9999{\longleftrightarrowfill@}{#1}{#2}}
\newcommand\longleftrightarrowfill@{%
  \arrowfill@\leftarrow\relbar\rightarrow}
\makeatother

\begin{document}
\thispagestyle{empty}
\vspace*{0.8cm} 
\begin{center}
{\Huge Tops as Building Blocks for $G_2$ Manifolds}\\

  \vspace*{2.5cm}
 Andreas P. Braun \footnote{e-mail: {\tt andreas.braun@maths.ox.ac.uk}} \\
 
  \vspace*{1.0cm} 
 {\it Mathematical Institute, University of Oxford, \\ Andrew Wiles Building, Woodstock Rd, Oxford OX2 6GG, UK }\\

\vspace*{0.8cm}
\end{center}
\vspace*{2.cm}

\noindent 

A large number of examples of compact $G_2$ manifolds, relevant to supersymmetric compactifications of M-Theory to four dimensions, can be constructed by forming a twisted connected sum of two appropriate building blocks times a circle. These building blocks, which are appropriate $K3$-fibred threefolds, are shown to have a natural and elegant construction in terms of tops, which parallels the construction of Calabi-Yau manifolds via reflexive polytopes.

\newpage

\tableofcontents


\section{Introduction}

Given the central position of M-Theory in the web of string dualities, a better understanding of M-Theory compactifications is expected to tie together supersymmetric compactifications of all of the weakly coupled string theories. While this makes compactification of M-Theory on manifolds of $G_2$ holonomy one of the most interesting parts of the landscape of string vacua, the number of concrete examples studied is relatively sparse as compared to compactifications of type II String Theories, heterotic String Theory, or F-Theory on Calabi-Yau manifolds. The main reason for the lack of examples is the comparatively easy construction of Calabi-Yau manifolds, which, using Yau's theorem, benefits from a host of techniques in complex algebraic geometry. The construction of $G_2$ manifolds as twisted connected sums \cite{corti_weakfano,corti_g2} allows such techniques to be applied
in the study of $G_2$ manifolds as well. It is the motivation of the present work to develop further techniques in this direction.

Examples of manifolds with metrics of $G_2$ holonomy fall into three broad classes. The first constructions \cite{bryant1989,gibbons1990} were non-compact examples which are asymptotically conical, see also \cite{Cvetic:2001zx,Cvetic:2001ih} for more examples. These can be thought of as smoothed versions of singular cones, which is very interesting from the point of view of physics: in compactifications of M-Theory on manifolds of $G_2$ holonomy, non-abelian gauge groups and (chiral) matter arise from singularities of the compactification geometry and interesting singularities can be constructed from such conical manifolds.

The first compact examples of $G_2$ manifolds are the famous Joyce manifolds \cite{joyce2000compact}. They are constructed from smoothings of toroidal orbifolds $T^7/G$. Unfortunately, (the singular versions of) these manifolds do not support interesting singularities on their own and their toroidal origin renders them locally flat. This prevents to simply cut and paste the interesting singular non-compact examples studied in the literature as these are not asymptotically flat.

A third class, which also contains compact $G_2$ manifolds, will be the subject of the present work. As first discussed by Kovalev \cite{kovalev00} and further elaborated on in \cite{corti_weakfano,corti_g2}, one may construct compact manifolds of $G_2$ holonomy by forming a twisted connected sum (TCS) of two appropriate `building blocks' times a circle. These building blocks can be thought of as $K3$ fibrations over a $\P^1$ base. Besides the two building blocks, the defining data of a TCS $G_2$ manifold includes the diffeomorphism used to glue two building blocks. On the level of the $K3$ fibres, this must be a hyper-K\"ahler rotation.

The simplest construction of building blocks was already contained in \cite{kovalev00}, where blowups of appropriate curves in Fano threefolds were employed. This has been generalized to so-called semi-Fano threefolds in \cite{corti_weakfano}. Another construction of building blocks is presented in \cite{Kovalev_Lee}. Here, non-symplectic involutions of $K3$ surfaces (classified in \cite{Nikulinaut1,Nikulinaut2,Nikulinaut3}) are used to construct building blocks by 
orbifolding $K3 \times \P^1$ and resolving singularities.

For a pair of building blocks (times a circle) to be glueable into a manifold of $G_2$ holonomy, one needs to find a diffeomorphism between the two $K3$ fibres which acts as a particular type of hyper-K\"ahler rotation, also called a `Donaldson matching'. Even though the conditions for this to exist can be spelt out purely in terms of lattice data (for lattice polarized families), it is not straightforward to determine in general if (and how many) such maps can be found. 

There are hundreds of thousands of semi-Fano threefolds, giving rise to many examples of building blocks. From the glueing techniques used in \cite{corti_g2}, it has been possible to mass produce $G_2$ manifolds and at least 50 million examples are known.

The present work starts from the observation that fibrations of toric hypersurface Calabi-Yau threefolds can be very easily detected in Batyrev's reflexive polytopes \cite{1993alg.geom.10003B,Klemm:1995tj,Hosono:1995bm,Avram:1996pj,Kreuzer:1997zg,Kreuzer:2000qv}. For a toric hypersurface Calabi-Yau threefold to be $K3$ fibred, the corresponding four-dimensional ${\bf N}$-lattice polytope $\Delta^\circ$ must contain a three-dimensional subpolytope $\Delta^\circ_F$ which also contains the origin as its unique interior point. The Picard lattice of the generic fibre is determined by $\Delta^\circ_F$. 
As it is codimension one, $\Delta^\circ_F$ furthermore cuts $\Delta^\circ$ into two halves (dubbed `top' and `bottom' in \cite{Candelas:1996su}) which contain information about degenerate fibres. Most reflexive four-dimensional polytopes give rise to $K3$ fibred Calabi-Yau threefolds and many of the structures found in the plot of the Hodge numbers based on the Kreuzer-Skarke database \cite{Kreuzer:2000xy} can be understood from this point of view \cite{Candelas:2012uu}.

While $K3$ fibred Calabi-Yau threefolds do not qualify as building blocks for $G_2$ manifolds, one may think of top and bottom as forming two halves into which a $K3$ fibred Calabi-Yau threefold can be disassembled. As shown in Section \ref{sect:tops_as_build}, which forms the 
core of this paper, it is possible to define a threefold purely from a top (or bottom). This threefold then has all of the properties of a building block as defined by \cite{kovalev00,corti_weakfano}. In particular, the fibre is a lattice polarized $K3$ surface, described as a toric hypersurface. The construction can be understood in an ad hoc way, but may also be described by using the language of (refinements of) normal fans of polytopes. The latter allows to exploit a wealth of results, such as the powerful methods of \cite{DK}. 

Section \ref{sect:examples} contains a detailed discussion of several examples which are meant to showcase the features of building blocks constructed from tops. In particular, we use a top origination from the largest reflexive four-dimensional polytope to construct a building block with contributes $240$ two-forms to any $G_2$ manifold glued from it.

Appendix \ref{sect:background} contains a collection of background material on lattices, the relation between mirror symmetry for $K3$ surfaces and Batyrev's reflexive polytopes (which is slightly more subtle than in the case of higher-dimensional Calabi-Yau manifolds). Appendix \ref{app:hodge_strat} demonstrates the familiar usage of the stratification methods of \cite{DK} in the case of toric hypersurface Calabi-Yau threefolds. Appendix \ref{sect:k3stabledeg} discussed the enlightening example of the stable degeneration of K3 surfaces into two $dP_9$ surfaces, which serves as an inspiration for our construction, in detail. Some technical details regarding the proof of the absence of torsion in $H^3(Z,\mathbb{Z})$ for building blocks $Z$ constructed from tops is relegated to Appendix \ref{sect:polybrauer}.

Some background of construction of $G_2$ manifolds via twisted connected sums is contained in Sections \ref{sect:defsbuildbocks}. As the relevant theorems and constructions from \cite{kovalev00,corti_g2,corti_weakfano} are also nicely reviewed in the physics literature in \cite{Halverson:2014tya}, the discussion is limited to a minimum.

\subsection{Building Blocks}\label{sect:defsbuildbocks}

The construction of $G_2$ manifolds as twisted connected sums \cite{kovalev00,corti_g2,corti_weakfano} starts from $K3$-fibred threefolds $Z$ which are 
called \emph{building blocks}. In this section, we review some of the properties of building blocks. 
Following \cite{corti_g2, corti_weakfano}, see also \cite{kovalev00},
we call an algebraic K\"ahler threefold $Z$ a building block if the following conditions are met:

\begin{itemize}
 \item[i)] $Z$ has a projection 
\begin{equation*}
\begin{array}{ccc}
 S & \hookrightarrow & Z \\
 &&\downarrow_\pi \\
 && \P^1
\end{array}
\end{equation*}
the generic fibres of which are non-singular $K3$ surfaces. 
\item[ii)] The anticanonical class of $Z$ is primitive\footnote{This means that 
there is no line bundle $L$ such that $L^{\otimes n} =  [K_Z]$ 
for any $n > 1$.} and equal to the class of the fibre, $S$:
\begin{equation*}
 [-K_Z] = [S]
\end{equation*}
\item[iii)] Picking a smooth and irreducible fibre $S_0$, we have a natural restriction map 
\begin{equation}\label{eq:defN}
\rho: H^2(Z,\mathbb{Z}) \rightarrow H^2(S_0,\mathbb{Z}) \cong \Lambda = (-E_8^{\oplus 2}) \oplus U^{\oplus 3}
\end{equation}
and there is no monodromy upon orbiting around $S_0$, i.e. the fibration is trivial in the vicinity of $S_0$.
Denoting the image of $\rho$ by $N$, we demand that the quotient $\Lambda/N$ is torsion free, i.e. the embedding $N 
\hookrightarrow \Lambda$ is primitive.
\item[iv)] $H^3(Z,\mathbb{Z})$ has no torsion\footnote{In fact, this is not strictly necessary, but simplifies the construction and the computation of the integral cohomology of the resulting $G_2$ manifolds.}. 
\end{itemize}
Under these assumptions, it follows that $Z$ is simply connected and the Hodge numbers $H^{1,0}(Z)$ and $H^{2,0}(Z)$ vanish. As $Z$ is a $K3$ fibration over
$\P^1$, the normal bundle of the fibre, and in particular of $S_0$, is trivial. The lattice $N$ naturally embeds into the Picard lattice of 
$S_0$ and we can think of the fibres as being elements of a family of lattice polarized K3 surfaces with polarizing lattice containing $L \supseteq N$. 

By excising a fibre, we may form the open space 
\begin{equation}
 V \equiv Z \setminus S_0 \, .
\end{equation}
from a building block $Z$. $V$ is an asymptotically cylindrical Calabi-Yau threefold. In particular, its Ricci-flat K\"ahler metric 
asymptotes to the Ricci-flat metric $ds^2$, K\"ahler form $\omega$ and holomorphic three-form 
$\Omega^{3,0}$ on the cylinder $\mathbb{R}_+ \times S^1 \times S_0$:
\be
\begin{aligned}
 ds^2 &= dt^2 + d\theta^2 + ds^2_{S_0} \\
\omega &= dt \wedge d\theta + \omega_{S_0} && \Omega^{3,0} = (d\theta - i 
dt)\wedge \Omega^{2,0}_{S_0} 
\end{aligned}
\ee
Here, $ds^2_{S_0}$, $\omega_{S_0}$ and $\Omega_{S_0}$ are the Ricci-flat metric, K\"ahler form 
and holomorphic two-form of $S_0$, $\theta$ is a coordinate on the $S^1$ and $t$ is a coordinate on $\R_+$.

\subsection{Gluing Building Blocks to \texorpdfstring{$G_2$}{Lg} Manifolds}\label{sect:matchingproblem}

Given two building blocks $Z_+$ and $Z_-$, we may glue $V_+ \times S^1$ and 
$V_- \times S^1$ along their asymptotic Calabi-Yau cylinder by swapping\footnote{There exists a more general `extra-twisted' version described in \cite{1505.02734}.} the extra $S^1$ with the $S^1$s of the cylinder while glueing the $K3$ surfaces with a diffeomorphism $g:S_{0+} \rightarrow S_{0-}$ acting as:
\be\label{eq:hkrot}
\begin{aligned}
g^*: \omega_{S_{0\pm}} &\leftrightarrow Re(\Omega_{S_{0\mp}}) \\
g^*: Im(\Omega_{S_{0\pm}}) &\leftrightarrow - Im(\Omega_{S_{0\mp}}) \, .
\end{aligned}
\ee
This realizes a manifold $X$ as a twisted connected sum which has a metric of $G_2$ holonomy \cite{kovalev00, corti_g2}. This metric is not the same as the Calabi-Yau metrics of the asymptotically cylindrical Calabi-Yau manifolds $V_{\pm}$ but it is, in a sense, close to these.

Note that one may think of $X$ as a (non-holomorphic) K3 fibration over a base $B$. This base $B$ is a fibration of a torus over an interval 
for which one of the two circles of the torus collapses at each end. This space is topologically a 3-sphere\footnote{I'd like to thank Johannes Walcher 
for pointing this out to me.} which is made manifest by using Hopf coordinates on $S^3$.

By the Torelli theorem, there is a unique diffeomorphism inducing 
\eqref{eq:hkrot} if and only if \eqref{eq:hkrot} is the effect of a lattice isometry
\begin{equation}
g_\Lambda: H^2(S_{0+},\mathbb{Z}) \rightarrow  H^2(S_{0-},\mathbb{Z})  \, .
\end{equation}
Choosing a marking $h: \Lambda \cong H^2(S,\mathbb{Z})$ on one of the K3 surfaces, such an isometry defines primitive embeddings of $N_\pm \hookrightarrow \Lambda$. Let us denote the orthogonal complement of $N_\pm$ in $\Lambda$ by $T_\pm$. The cohomology groups of the resulting $G_2$ manifolds $X$ are then given by \cite{corti_g2}:
\begin{equation}\label{eq:bettiG2}
\begin{aligned}
H^1(X,\mathbb{Z}) & =   0 \\
H^2(X,\mathbb{Z}) & =  N_+ \cap N_- \oplus K_+ \oplus K_- \\
H^3(X,\mathbb{Z}) & = \mathbb{Z}[S] \oplus \Lambda/(N_+ + N_-) \oplus (N_- \cap T_+) \oplus (N_+ \cap T_-)\\
& \hspace{1cm} \oplus H^3(Z_+)\oplus H^3(Z_-) \oplus K_+ \oplus K_- \\
H^4(X,\mathbb{Z}) & = H^4(S) \oplus (T_+ \cap T_-) \oplus \ \Lambda/(N_- + T_+) \oplus \Lambda/(N_+ + T_-) \\
& \hspace{1cm} \oplus  H^3(Z_+)\oplus H^3(Z_-) \oplus K_+^* \oplus K_-^* 
\end{aligned}
\end{equation}
Here, the group $K$ is defined as
\begin{equation}
K \equiv \mbox{ker} (\rho)/[S_0] \, .
 \end{equation}
and $K^*$ is its dual.

\section{Tops as Building Blocks} \label{sect:tops_as_build}

In this section we introduce the main idea of this paper and show how 
appropriate, `projecting', tops can be used to construct building blocks as 
defined above. The necessary background in toric geometry can be
found e.g. in \cite{Kreuzer:2006ax, Fulton, danilov}.

\subsection{Tops and Fibred Calabi-Yau Manifolds}

In its most general incarnation \cite{Bouchard:2003bu} a top $\Diamond^\circ$ 
is defined as a bounded lattice polytope (with respect to a lattice ${\bf N}$) 
satisfying relations of the form
\be
\begin{aligned}\label{eq:topsbouchard}
\langle m_i, \Diamond^\circ &\rangle \geq -1 \\
\langle m_0, \Diamond^\circ &\rangle \geq 0 
\end{aligned}
\ee
for a set of lattice points $m_i$ and a single $m_0$, all sitting in the dual lattice ${\bf M}$. 

Although not the most general case, thinking of a top as half a reflexive 
polytope \cite{1993alg.geom.10003B}, $\Delta^\circ$ in the ${\bf N}$-lattice, 
is sufficient for our purposes. This is also how the concept originally 
appeared in \cite{Candelas:1996su}. A reflexive lattice polytope $\Delta^\circ$, is a convex polytope such that its dual $\Delta$ in ${\bf M}$, defined by  
\begin{equation}\label{eq:dualpolytopes}
 \langle \Delta, \Delta^\circ \rangle \geq -1 \, ,
\end{equation}
is also a lattice polytope. Such a pair of lattice polytopes defines a family of Calabi-Yau hypersurfaces as follows. A triangulation of the polytope $\Delta^\circ$ defines a fan $\Sigma$ by taking all cones over the simplices of the triangulation on the surface of $\Delta^\circ$ \footnote{The condition \eqref{eq:dualpolytopes} guarantees that the origin is the unique interior point of $\Delta^\circ$.}. This in turn defines a toric variety $\P_\Sigma$ which is projective, i.e. has a non-zero K\"ahler cone, if the associated star triangulation, for which all simplices contain the origin as a vertex, is regular \cite{Fulton,rambau}. The dual polytope $\Delta$ is the Newton polytope of a generic section $P$ of $-K_{\P_\Sigma}$ and we may find a family of Calabi-Yau hypersurfaces as 
\begin{equation}\label{eq:CYfrompolytopes}
  P = \sum_m c_m \prod_{\nu_i} z_i^{\langle m, \nu_i  \rangle + 1} = 0  \, .
\end{equation}
Here $\nu_i$ are lattice points on $\Delta^\circ$, which are associated with rays of the fan $\Sigma$ and hence with homogeneous coordinates $z_i$ of $\P_\Sigma$. The sum runs over lattice points on $\Delta$ and the $c_m$ are complex coefficients. If the polytopes $\Delta$ and $\Delta^\circ$, i.e. the lattices ${\bf M}$ and ${\bf N}$ are $n-$dimensional, the toric variety $\P_\Sigma$ is complex $n-$dimensional and \eqref{eq:CYfrompolytopes} gives a complex $n-1$ dimensional Calabi-Yau manifold $X_{(\Delta,\Delta^\circ)}$. For a sufficiently generic choice of $c_m$, there always exists a triangulation such that above always gives a smooth Calabi-Yau manifold if $n\leq 4$. Exchanging the roles of $\Delta$ and $\Delta^\circ$ gives a construction of the mirror family of Calabi-Yau manifolds.

Starting from $\Delta$, the polytope $\Delta^\circ$ is uniquely determined (via \eqref{eq:dualpolytopes}) by the vertices of $\Delta$. This gives the first relation of \eqref{eq:topsbouchard}. From this point of view, the second relation of \eqref{eq:topsbouchard} cuts $\Delta^\circ$ through the origin and leaves us with only one of its halves. Cutting polytopes $\Delta^\circ$ in this fashion reveals fibration structures of $X_{(\Delta,\Delta^\circ)}$ by Calabi-Yau manifolds of one lower dimension \cite{Klemm:1995tj,Hosono:1995bm,Avram:1996pj,Kreuzer:1997zg,Kreuzer:2000qv}, as we will review in the following. 

Before specializing to our case of interest, which will be Calabi-Yau threefolds fibred by $K3$ surfaces, let us give the general picture of how fibration structures can be found using polytopes. Let us assume that we can find a subspace $F$ of ${\bf N}\otimes\R$ such that $\Delta^\circ_F \equiv \Delta^\circ \cap F$ is again a reflexive lattice polytope in ${\bf N}_f = {\bf N} \cap F$. In this case there are the dual short exact sequences 
\begin{equation}
 \begin{aligned}
  0 \hspace{.3cm} \rightarrow &\hspace{.3cm} {\bf N}_f& \rightarrow &\hspace{.3cm}{\bf N}& \rightarrow &\hspace{.3cm}{\bf N}_b &\rightarrow \hspace{.3cm} 0 \\
  0 \hspace{.3cm} \rightarrow &\hspace{.3cm} {\bf M}_b& \rightarrow &\hspace{.3cm}{\bf M}& \rightarrow &\hspace{.3cm}{\bf M}_f &\rightarrow \hspace{.3cm} 0 
 \end{aligned}
\end{equation}
defining lattices ${\bf N}_b$ and ${\bf M}_b$. The projection to ${\bf N} \rightarrow {\bf N}_b$ gives rise to a projection morphism on the level of the ambient space if the induced map on the fan $\Sigma \rightarrow \Sigma_b$ is a toric morphism, i.e. maps each cone of $\Sigma$ into a cone of $\Sigma_b$ (see \cite{Fulton} Section 1.3). The fibres of this projection applied to the Calabi-Yau hypersurface $X_{(\Delta,\Delta^\circ)}$ are again Calabi-Yau hypersurfaces from the algebraic family $X_{(\Delta_F,\Delta^\circ_F)}$. The Calabi-Yau hypersurface $X \subset \mathbb{P}_\Sigma$ hence enjoys a fibration by another Calabi-Yau manifold $X_F$ of lower dimension which is determined by the polytope $\Delta^\circ_F$ 
\cite{Klemm:1995tj,Hosono:1995bm,Avram:1996pj,Kreuzer:1997zg,Kreuzer:2000qv}.

We now specialize to the case where $F$ (and, correspondingly, $X_F$) is a codimension one submanifold. In this case, we may write $F = m_0^\perp$
for a lattice vector $m_0$ and $F$ separates $\Delta^\circ$ into the two halves
\begin{equation}
 \begin{aligned}
  \Diamond_1^\circ \equiv \{\nu \in \Delta^\circ | \langle \nu,m_0 \rangle \geq 0 \} \\
  \Diamond_2^\circ \equiv \{\nu \in \Delta^\circ | \langle \nu,m_0 \rangle \leq 0 \} 
\end{aligned}
\end{equation}
which were called \emph{top} and \emph{bottom} in \cite{Candelas:1996su}. 

To see the fibration structure explicitely, it is convenient to introduce an equivalence relation on the set of all lattice points of $\Delta$ as follows 
\begin{equation}
m \sim m' \,\,\,\,\mbox{if} \,\,\,\,m - m' = k m_0. 
\end{equation}
for an integer $k$.
Denoting the set of equivalence classes under $\sim$ by $\mathcal{M} = \{[M]\}$,  the defining polynomial of a generic hypersurface can be written as
\begin{equation}
\begin{aligned}
  & \sum_m c_m \prod_{\nu_i} z_i^{\langle m, \nu_i  \rangle + 1} \\
= & \sum_{[M]\in \mathcal{M}}\,\, \sum_{m \in [M]} c_m \prod_{\nu_i} z_i^{\langle m,\nu_i \rangle + 1 } \\
= & \sum_{[M]\in \mathcal{M}} \left(\prod_{\nu_i \in \Delta^\circ_F} z_i^{\langle m,\nu_i \rangle + 1 } \right)
\underbrace{\left( \sum_{m \in [M]} c_m \prod_{\nu_i \notin \Delta^\circ_F} z_i^{\langle m, \nu_i \rangle + 1} \right)}_{c_m^F}
\end{aligned}
\end{equation}
i.e. we have a hypersurface equation of a Calabi-Yau manifold $X_F$ determined by $\Delta^\circ_F$,
the coefficients $c_m^F$ of which are dependent on the remaining coordinates, as expected for a fibration.

For Calabi-Yau manifolds fibred by Calabi-Yau manifolds of one lower dimension, the base of the fibration must be a $\P^1$. We can write the homogeneous coordinates $[z_t:z_b]$ of this $\P^1$ explicitely as 
\begin{equation}\label{eq:coordsbase}
\begin{aligned}
 z_t &= \prod_{\nu_i| \nu_i \in \Diamond_1^\circ } z_i^{\langle m_0,\nu_i \rangle} \\
 z_b &= \prod_{\nu_i| \nu_i \in \Diamond_2^\circ } z_i^{-\langle m_0,\nu_i \rangle} 
\end{aligned}
\end{equation}
Note that coordinates associated with $\nu_i \in \Delta^\circ_F$ do not contribute because $\langle  m_0 ,\nu_i\rangle = 0$. Furthermore, note that each sum only runs over lattice points above or below $F$. The class of the fibre can be found by fixing a generic point on the base $\P^1$. 
From \eqref{eq:coordsbase} it follows that the toric divisors classes $[D_i] = [\{z_i=0\}]$ obey 
\begin{equation}\label{eq:topfibreclass}
 [X_F] = \sum_{\nu_i \in \Diamond^\circ_1} \langle \nu_i, m_0 \rangle D_i =  - 
\sum_{\nu_j \in \Diamond^\circ_2} \langle \nu_j, m_0 \rangle D_j
\end{equation}
Points above (below) F hence correspond to fibre components over $z_t=0$ and $z_b$ = 0. In case a top only has a single point above $F$, the fibre stays irreducible over $z_t=0$. 

As we have seen, lattice points on the top and the bottom contain information about degenerate fibres\footnote{Additionally, lattice points interior to faces of maximal dimension on the polytope characterizing the fibre give rise to additional degenerate fibres if they are not contained in faces of maximal dimension of $\Delta^\circ$.}. As such they may not only be used to study fibrations by lower-dimensional Calabi-Yau manifolds, but can also be thought of as characterizing degenerations of the fibre. See \cite{davis_tops} for the relation of tops to semi-stable degenerations. This correspondence is particularly nice in the case of three-dimensional tops, corresponding to elliptically fibred K3 surfaces, which show the extended Dynkin diagrams of the corresponding gauge groups in F-theory \cite{Candelas:1996su,Candelas:1997pv,Perevalov:1997vw}. See also \cite{Borchmann:2013jwa,Borchmann:2013hta} for an elegant application to models with non-trivial abelian gauge sector in F-theory.

\subsection{Ad hoc Construction, Projecting Tops and Elementary Properties}\label{sect:adhocconstruction}

In this work, we use the term top for any lattice polytope $\Diamond^\circ$ which can be described as
\begin{equation}
 \Diamond^\circ = \{\nu \in \Delta^\circ | \langle \nu,m_0 \rangle \geq 0 \} \, 
\end{equation}
for a reflexive polytope $\Delta^\circ$ and a vector $m_0$ such that 
\begin{equation}
 \Delta_F^\circ = \{\nu \in \Delta^\circ | \langle \nu,m_0 \rangle = 0 \} = \Diamond^\circ \cap F
\end{equation}
is again reflexive. Furthermore, we are interested in four-dimensional tops, for which 
$\Delta_F^\circ = \Diamond^\circ \cap F$ is a three-dimensional reflexive polytope.
In many case, two such tops which share the same $\Delta_F^\circ$ may be combined to define a reflexive polytope $\Delta^\circ$ and hence a $K3$-fibred Calabi-Yau threefold and it seems intuitively clear that we should be able to cut the base $\P^1$ of such a $K3$ fibred
Calabi-Yau manifold $X$ into two halves, each half corresponding to a single top. 
In fact, this is a well-known story in the case that $X$ is an 
elliptic K3 surface: here, tops whose projection to $F$ lies on $\Delta^\circ_F$
correspond to families of rational elliptic surfaces (also called $dP_9$ in 
the physics literature) and the points on the top correspond to the components 
of degenerate fibres of the elliptic fibration. Decomposing a 
K3 surface into two rational elliptic surfaces is accomplished by means of a 
specific stable degeneration limit, reviewed in Appendix \ref{sect:k3stabledeg}.
Note that we can think of $dP_9$ surfaces as being an analogue of the building blocks defined in Section \ref{sect:defsbuildbocks} in one dimension lower. They are complex surfaces fibred by Calabi-Yau one-fold (elliptic curves), the first Chern class of which is represented by the class of the fibre. Here, we are not interested in the higher-dimensional version of this stable degeneration limit, but we want 
to directly describe the final products, i.e. the components of the central fibre 
of the degeneration limit.

Before we proceed, we will need to make a further technical assumption, which
should become clear in the following. For now, let us merely 
motivate it by the following observation: as mentioned above, only specific tops are associated with the degeneration of a $K3$ surface into two $dP_9$ surfaces\footnote{Note that the degeneration of a $K3$ surface 
used to study the duality to heterotic $SO(32)$-string theory is also associated with tops, but does not lead to 
a pair of rational elliptic surfaces but rather degenerates every
fibre of an elliptic $K3$ surface \cite{Aspinwall:1997ye}, similar to the 
realization of the Sen limit via stable degeneration \cite{Clingher:2012rg}.}. Those tops are such that there projection onto $F$ lies on $\Delta_F^\circ$.

Following \cite{Candelas:2012uu}, we call tops $\Diamond^\circ$ for which the projection 
onto $F$ is contained in $\Delta_F^\circ$ \emph{projecting}. In the rest of this paper, we shall assume that the tops we are working with all have this property. 

An interesting property of such tops is that we may form a reflexive polytope from any pair of projecting tops $\Diamond^\circ$ and $\Diamond^{\circ'}$ for which $\Delta_F^\circ = \Delta_F^{\circ'}$. For this reason, they were used in \cite{Candelas:2012uu} to understand the intricate structures found in the plot of Hodge numbers of toric hypersurface Calabi-Yau threefolds.

Forming a reflexive polytope from two tops sharing a common $\Delta^\circ_F$, the dual polytope $\Delta$ also has a subpolytope given by $\Delta_F$, which is the polar dual of $\Delta^\circ_F$ \cite{Avram:1996pj}. Hence in this case both a Calabi-Yau threefold and its mirror are $K3$ fibred by $K3$ surfaces from algebraic mirror families. Note that this does not mean that mirror symmetry acts fibre-wise on the $K3$ surfaces.

Note that we may exploit $SL(4,\mathbb{Z})$ to put $m_0$ in the convenient location $(0,0,0,1)$, which is what we assume in the following. 
This means that all points on $\Delta_F^\circ$ have the form $(\vec{v}_F,0)$. As we are only interested in projecting tops, all other points on $\Diamond^\circ$ have the form $(\vec{v},h)$ for $h > 0$ and 
$\vec{v} \in \Delta^\circ_F$.

We are now ready to present our construction.
In the same way that a polytope $\Delta^\circ$ defines a toric variety by taking the cones over all faces of $\Delta^\circ$ and then resolving singularities by refining cones according to a triangulation, we may construct a toric variety from $\Diamond^\circ \cup \nu_0$. We denote the generators of the rays of the resulting fan (i.e. lattice points on $\Diamond$) by $\nu_i$, with $\nu_0 = (0,0,0,-1)$. In a sense, this means we are turning $\Diamond^\circ$ into a reflexive polytope by completing it with the minimal bottom. Let us denote the fan constructed from a fine regular triangulation of $\Diamond^\circ \cup \nu_0$ by $\Sigma$ and the corresponding toric variety by $\mathbb{P}_\Sigma$. This completes the construction of the ambient space.

The hypersurface equation of our building block is now given by taking the set of 
all points $\Diamond$ in the ${\bf M}$-lattice with the property
\be
\begin{aligned}\label{eq:conddualtop}
\langle \Diamond, \nu_i \rangle \geq -1 \\
\langle \Diamond, \nu_0  \rangle \geq 0 \, . 
\end{aligned} 
\ee
To each such point $m$ we associate the monomial
\begin{equation}\label{eq:topmonos}
z_0^{\langle m , \nu_0\rangle}  \prod_{\nu_i} z_i^{\langle m, \nu_i \rangle+1} \, .
\end{equation}
We may then sum up all such monomials with generic coefficients $c_m$ to find a homogeneous hypersurface equation for a hypersurface $Z$ in $\mathbb{P}_\Sigma$. The points satisfying \eqref{eq:conddualtop} naturally define a top in the ${\bf M}$-lattice which is why we denote it by $\Diamond$ and call it the dual top. Note that $\Diamond$ and $\Diamond^\circ$ are not polar dual polytopes and that our definition of dual top is not the same as that adapted in \cite{Bouchard:2003bu}. Rather, $\Diamond$ is obtained from the polar dual of $\Diamond^\circ\cap \nu_0$ by excising all points for which $\langle \Diamond, \nu_0 \rangle < 0$. 

What we have done closely mimics the result of the procedure outlined in Appendix \ref{sect:k3stabledeg}: the new coordinate $z_0$ corresponds to $\lambda_1$ and the condition $\langle m, \nu_0 \rangle \geq 0$ eliminates all monomials which contain non-zero powers of $\lambda_2$.

As we may combine any two projecting tops $\Diamond^\circ_1$ and $\Diamond^\circ_2$ with the same $\Delta^\circ_F$ to form a reflexive polytope, we may in particular use $\Diamond^\circ$ along with a copy ${\Diamond^\circ}'$ of itself (with the fourth coordinate inverted). 
The polar dual $\Delta$ of $\Delta^\circ = \Diamond^\circ \cup {\Diamond^\circ}'$ is then the same as
$\Delta = \Diamond \cup {\Diamond}'$ with $\Diamond$ given by \eqref{eq:conddualtop}. Using the result of
\cite{Avram:1996pj} cited above it follows that $\Diamond \cap F = \Delta_F$, where  $\Delta_F$ is the polar dual
of $\Delta^\circ_F$ in $F$.

A dual top $\Diamond$ is also projecting if $\Diamond^\circ$ has this property. As $\Diamond^\circ$ 
sits above $F$ and $\Diamond$ below $F$, the product of the fourth components for each point $\nu \in \Diamond^\circ $ and $m\in \Diamond$ is always non-positive. Hence
\begin{equation}
 -1 \leq \langle m,\nu \rangle =  \langle m|_F, \nu|_F \rangle + \nu_4 \cdot m_4 \leq  \langle m|_F, \nu|_F\rangle \, .
\end{equation}
As we have that $\Diamond^\circ$ is projecting, $\nu|_F$ is contained in $\Delta^\circ_F$. The above relation then
forces $ m|_F$ to be contained in $\Delta_F$, so that $\Diamond$ must be projecting as well.

Repeating the analysis of the last section, we find that $Z$ is a $K3$-fibred threefold
with base $\P^1$. The homogeneous coordinates of the $\P^1$ base are
\begin{equation}
 [z_0:\prod_{\nu_i} z_i^{\langle m_0, \nu_i \rangle}]\, ,
\end{equation}
and the class of the fibre $X_F$ is given by
\begin{equation}\label{eq:fibrecomp}
 [X_F] = [z_0] = \sum_i  \langle m_0, \nu_i \rangle D_i  \, .
\end{equation}
We immediately find by adjunction that 
\begin{equation}
 -[K_Z] =  \left( D_0 + \sum_i D_i \right) -\sum_i D_i = D_0 = [X_F]\, , 
\end{equation}
i.e. the anticanonical class of $Z$ equals that of the fibre. 

We can write the defining polynomial of a generic hypersurface as
\begin{equation}\label{eq:Zpoly}
\sum_{[M]} \underbrace{\left( \sum_{m \in [M]} c_m \,\, z_0^{\langle\nu_0,m \rangle}  \prod_{\nu_i \notin \Delta^\circ_F} z_i^{\langle \nu_i , m \rangle + 1} \right)}_{c_m^F} \left(\prod_{\nu_i \in \Delta^\circ_F} z_i^{\langle \nu_i , m \rangle + 1 } \right) \, .
\end{equation}
We associate the fibre over the point $z_0 = 0$ with $S_0$. It has a particularly simple description: first note that setting $z_0 = 0$ we cut out a toric variety $\P_{\Sigma_F}$ from the ambient space $\P_\Sigma$. As $\nu_0$ never shares a cone with any of the $\nu_i$ not on $\Delta^\circ_F$, we can effectively set all of the corresponding 
coordinates to $1$. Said more abstractly, $\Sigma_F = \mbox{star}(\nu_0)$ is a fan over the faces of $\Delta^\circ_F$. On the level of \eqref{eq:Zpoly},
setting $z_0 = 0$ eliminates all terms except the ones for which $\langle \nu_0, m \rangle = 0$. As these are precisely the lattice points of $\Diamond$
on $F$, we find that the fibre over $z_0 = 0$ is a $K3$ surface $S$ defined in terms of the pair of reflexive polytopes $\Delta^\circ_F , \Delta_F$, described
in Section \ref{sect:mirk3}.

We can describe a generic fibre by intersecting \eqref{eq:Zpoly} with 
\begin{equation} \label{eq:genericfibre}
\alpha z_0 = \beta \prod_{\nu_i} z_i^{\langle m_0 ,\nu_i\rangle}
\end{equation}
for two complex numbers $\alpha$ and $\beta$. Away from $\alpha = 0$ only lattice points on $\Delta^\circ_F$ meet the $K3$ fibre. Hence 
all fibres except the (potentially) reducible ones are from the family of lattice polarized $K3$ surfaces with lattice polarization $L = Pic(S)$.
For a non-trivial top, there is a reducible fibre with components $\prod_{\nu_i} z_i^{\langle \nu_i,m_0 \rangle}=0$. There can be additional reducible fibres if $\Delta^\circ_F$ contains lattice points interior to its two-dimensional faces, see Section \ref{sect:sexticfibre} for an example.

To show that the first two requirements of \cite{corti_g2}, i.e. i) and ii) from Section \ref{sect:defsbuildbocks}, are satisfied, we still need to show that the embedding of $S$ into $H^2(Z,\mathbb{Z})$ is primitive. If $S = [z_0]$ were not primitive, there must be a reducible fibre with several components all giving rise to equal classes in cohomology. Hence $\prod_{\nu_i} z_i^{\langle m_0 ,\nu_i\rangle}$ must be equal
to $n D$ for some divisor $D$. We now argue that this is impossible. First note that in case there are several $\nu_i$ such that $\langle m_0, \nu_i \rangle > 0$, these cannot be all equivalent (or a multiple over $\Q$ of some class) as there cannot be a corresponding linear relation. Furthermore, for any projecting top, $\langle m_0, \nu_i \rangle > 1$ implies that there is at least a second $\nu_j$ with $\langle m_0, \nu_j \rangle = 1$. Hence $[z_0] \neq n D_i$ for $ n >1 $ for any $D_i$ inherited from the ambient space. If some of the $D_i$ contributing to \eqref{eq:genericfibre} are reducible on $Z$, each extra components will give rise to an independent class in $H^{1,1}(Z)$ in accordance with \eqref{eq:hodgenumbersZ}, so that $[z_0]$ can never be a multiple of a single class in $H^2(Z)$. Reducible fibres resulting from interior points
of two-dimensional faces of $\Delta^\circ_F$ can be treated similarly.

\subsection{Formal Construction}\label{sect:formalconstruction}

In order to continue our analysis and prove all the requirements for $Z$ to be a building block, \eqref{sect:defsbuildbocks}, we need to introduce some more machinery. This will also explain why the ad hoc construction presented above gives rise to a threefold with all of the desired properties. Furthermore, it will allow us to give elegant combinatorial formulae for the Hodge numbers of $Z$ and the ranks of the lattices $N$ and $K$. 

For any polytope $\Delta$ the ${\bf M}$-lattice, there is an associated normal fan $\Sigma_n(\Delta)$ giving rise to a toric variety $\P_{\Sigma_n(\Delta)}$ along with a divisor $D_{\Delta}$. This description allows to use the theory of \cite{DK}, which is a very powerful tool to analyse the geometry of a hypersurface corresponding to $D_\Delta$ by means of stratification. 

The normal fan of a polytope is defined as follows: to every face $\Theta^{[k]}$ of a polytope $\Delta$, we may associate a cone
\begin{equation}
\check{\sigma}_n(\Theta^{[k]}) = \bigcup_{r \geq 0} r \cdot (p_\Delta - p_{\Theta^{[k]}})
\end{equation}
where $p_\Delta$ is an arbitrary point lying inside $\Delta$ and $p_\Theta^{[k]}$ 
is an arbitrary point lying inside $\Theta^{[k]}$. The dual cones $\sigma_n(\Theta^{[k]})$, defined by
\begin{equation}
\langle \check{\sigma}_n, \sigma_n \rangle \geq 0 \, ,
\end{equation}
form a complete fan which is called the \emph{normal fan} $\Sigma_n(\Delta)$ of $\Delta$. Here, $k$-dimensional faces $\Theta^{[k]}$ of $\Delta$
are associated with $4-k$-dimensional cones $\sigma_n(\Theta^{[k]})$ (for four-dimensional polytopes). 

On the normal fan, there is a convex support function $\Psi_\Delta$, linear on each cone of $\Sigma_n$. For each cone of maximal dimension (four in our case) $\Psi_\Delta$ can be described by
using the corresponding vertex $m_i$ and setting:
\begin{equation}
\left.\Psi_\Delta\right|_{\sigma_n(m_i)}  (p) =  \langle m_i, p \rangle
\end{equation}
for each point $p$ in $\sigma_n(m_i)$. This also determines $\Psi_\Delta$ for all cones of lower dimension.
The divisor 
\begin{equation} \label{eq:divnfD}
D_\Delta = \sum_{\nu_j \in  \Sigma_n(1)}  a_j D_j 
\end{equation}
can then be determined from
\begin{equation} \label{eq:supfct}
\left.\Psi_\Delta\right|_{\sigma_n(m_i)}  (\nu_j) = - a_j \,\,\, \forall \nu_j \in \sigma_n(m_i),
\end{equation}
and convexity means that
\begin{equation}
\left.\Psi_\Delta\right|_{\sigma_n(m_i)} (\nu_j) > - a_j \,\,\, \forall \nu_j \notin \sigma_n(m_i)
\end{equation}
In this construction, points of $\Delta$ are associated with holomorphic sections of $\mathcal{O}(D_\Delta)$ and $\Delta$ becomes the Newton polyhedron of
a generic hypersurface defined by the zero locus of a section of the line bundle $\mathcal{O}(D_\Delta)$. 

This construction is particularly simple in the case of reflexive pairs of polytopes. Here, the normal fan $\Sigma_n(\Delta)$ is equal to the fan over the faces $\Sigma_f(\Delta^\circ)$ of its polar dual and $D_\Delta = -K_{\P_{\Sigma_n(\Delta)}}$. In particular, for $n$-dimensional polytopes there is a one-to-one correspondence between $k$-dimensional faces of $\Delta$ and $n-k-1$-dimensional faces of $\Delta^\circ$. A (regular, fine, star) triangulation of $\Delta^\circ$ then defines a refinement of the normal fan giving rise to a (partial) crepant desingularization of a Calabi-Yau hypersurface. 

\subsubsection{Normal Fans of Tops}\label{sect:normalfanstops}

In the case at hand, we are interested in the normal fan $\Sigma_n(\Diamond)$ for the polytope $\Diamond$ defined in \eqref{eq:conddualtop}. Consider the face $\Theta^{[3]}_0 \equiv \Diamond \cap F$ of $\Diamond$. Here, we find that $\check{\sigma}_n(\Theta^{[3]}_0)$ is the entire lower half-space, bounded by $F$. Hence the dual cone appearing in the normal fan is
\begin{equation}
 \sigma_n(\Theta^{[3]}_0) = \mbox{cone}(\nu_0) \, .
\end{equation}
This is good news, as the ambient space used in the ad hoc construction presented above not only includes rays over lattice points of $\Diamond^\circ$, but also the extra vertex $\nu_0 = (0,0,0,-1)$, so that we get a compact toric variety. It is not hard to realize, however, that in general
\begin{equation}\label{eq:fvsnneq}
\Sigma_n(\Diamond) \neq \Sigma_f(\Diamond^\circ \cup \nu_0) \, ,
\end{equation}
where we use $\Diamond^\circ \cup \nu_0$ to denote the polytope which forms the convex hull over the union of $\Diamond^\circ$ and $\nu_0$. To see this, note that the polytope $\Diamond^\circ \cap \nu_0$ is in fact reflexive for any projecting top $\Diamond^\circ$. Hence its face fan $\Sigma_f$ is equal to the normal fan of its polar dual, which can be described as the convex hull of 
\begin{equation}
 \Diamond \cup (\Delta_F,1) \, .
\end{equation}
As the normal fans of $\Diamond$ and $\Diamond \cup (\Delta_F,1)$ are, in general, very different, \eqref{eq:fvsnneq} follows. 

As discussed in the following, a weaker statement can be shown, however. Namely, $ \Sigma_f(\Diamond^\circ \cup \nu_0)$ is a refinement 
\begin{equation}
  \Sigma_f(\Diamond^\circ \cup \nu_0) \rightarrow \Sigma_n(\Diamond) \, 
\end{equation}
of $\Sigma_n(\Diamond)$. 

%

In order to describe the normal fan of a top $\Diamond$ and its relation to the dual top $\Diamond^\circ$ defined by \eqref{eq:topsduality}, let us try to exploit the known relations between a pair of reflexive polytopes. First of all, $\Diamond^\circ \cap F = \Delta_F^\circ$ and $\Diamond \cap F = \Delta_F$ form a reflexive pair of three-dimensional polytopes defining the algebraic family of the $K3$ fibre. 
We may extend $\Diamond^\circ$ to a polytope $\Delta^\circ$ by considering the convex hull of $\Diamond^\circ \cup \nu_0$. 
\begin{equation}\label{eq:defDeltacirc}
 \Delta^\circ = \Diamond^\circ \cup \nu_0 \, .
\end{equation}
$\Delta^\circ$ is reflexive by the results of \cite{Avram:1996pj,Candelas:2012uu} as it is built from two projecting tops sharing the same $\Delta_F^\circ$. The dual polytope $\Delta$ is described as the convex hull of $\Diamond$ together with $\Delta_F$ displaced by $m_0$:
\begin{equation}\label{eq:defDelta}
 \Delta = \Diamond \cup (\Delta_F, 1) \, .
\end{equation}
As these polytopes form a reflexive pair, we have the usual relation 
\begin{equation}
\Sigma_f(\Delta^\circ) = \Sigma_n(\Delta) \, ,
\end{equation}
where 
\begin{equation}
 \sigma_f(\Theta^{\circ [k]}) = \sigma_n(\Theta^{[3-k]}) \, 
\end{equation}
for a dual pair of faces on $\Delta^\circ$ and $\Delta$ satisfying
\begin{equation}
 \langle (\Theta^{\circ [k]}, \Theta^{[3-k]} \rangle = -1 \, .
\end{equation}

Trivially, the set of faces of $\Delta^\circ$ is the same as the set of faces of $\Diamond^\circ \cup \nu_0$. However, the set of faces
of $\Delta$ is different from those of $\Diamond$. Consequently, not every cone $\sigma_f(\Theta^{\circ [k]})$ appears in $\Sigma_n(\Diamond)$. By construction, \eqref{eq:defDelta}, $\Delta$ contains precisely one face $\Theta^{[k+1]}$ for every face $\Theta_F^{[k]}$ which extends vertically upwards from $\Delta_F$. This means that such faces $\Theta^{[k+1]}$ are orthogonal to the plane $F$. Depending on $\Diamond$, $\Theta^{[k+1]}$ may extend also below $\Theta_F^{[k]}$, in which case $\Theta_F^{[k]}$ is a face of $\Diamond$, but not of $\Delta$. Under the polar duality of $(\Delta_F,\Delta_F^\circ)$, the face $\Theta_F^{[k]}$ is associated with a face $\Theta_F^{\circ [2-k]}$ on $\Delta_F^\circ$, so that
\begin{equation}
\langle \Theta_F^{[k]} ,\Theta_F^{\circ [2-k]} \rangle  = -1 \, .
\end{equation}
Hence polar duality on $(\Delta, \Delta^\circ)$ will identify $\Theta_F^{\circ [2-k]}$ with a face on $\Delta$ of dimension $k+1$ containing 
$\Theta_F^{[k]}$. As $\Theta^{[k+1]} \supset \Theta_F^{[k]}$ satisfies
\begin{equation}
 \langle \Theta^{[k+1]}, \Theta_F^{\circ [2-k]} \rangle = -1 \, ,
\end{equation}
we can identify $\Theta^{[k+1]}$ as this face and conclude that
\begin{equation}
\sigma_f(\Theta_F^{\circ [2-k]} ) = \sigma_n(\Theta^{[k+1]}) \, .
\end{equation}
When we collapse $\Delta$ to $\Diamond$, it can happen that a face $\Theta^{[k+1]}$ entirely disappears along with its cone in the normal fan. In this case there is no cone in $\Sigma_n(\Diamond)$ which is the cone over the face $\Theta_F^{\circ [2-k]}$. 

\subsubsection{Vertical and Non-Vertical Faces}\label{sect:vertnvert}

To describe in detail in which way $\Sigma_n(\Diamond)$ differs from $\Sigma_f(\Diamond^\circ)$, let us introduce the following definitions. A face $\Theta^{\circ [k+1]}$ ($\Theta^{[k+1]}$) of $\Diamond^\circ$ ($\Diamond$) is called vertical if it is orthogonal to $F$ and it is called non-vertical otherwise. A face $\Theta_F^{\circ [k]}$ ($\Theta_F^{[k+1]}$)
of $\Delta_F^\circ$ ($\Delta_F$) is called vertically embedded if there is a vertical face $\Theta^{\circ [k+1]}$ ($\Theta^{[k+1]}$) on $\Diamond^\circ$ ($\Diamond$) containing $\Theta_F^{\circ [k]}$ ($\Theta_F^{[k+1]}$) in its boundary. Such faces are called non-vertically embedded otherwise. As we are only considering projecting tops, it follows that for vertically embedded faces there exists a unique face $\Theta_F^{\circ [k+1]}\supset \Theta_F^{\circ [k]}$ ($\Theta_F^{[k+1]}\supset \Theta_F^{[k]}$) which extends upwards from $\Theta_F^{\circ [k]}$ (downwards from $\Theta_F^{[k]}$). This face must of course be vertical. Any other face above $F$ (below $F$) containing $\Theta_F^{\circ [k]}$ ($\Theta_F^{[k]}$) must be non-vertical and will hence be contained in the interior of $\Diamond^\circ$ ($\Diamond$) or the relative interior of one of its faces. For non-vertically embedded faces, however, there can be several non-vertical faces containing them in their boundary. 

We now observe that for a dual pair of faces $\Theta_F^{\circ [k]}$, $\Theta_F^{[2-k]}$, it will always be the case that one is vertically embedded, whereas the other is non-vertically embedded. First note that by $\langle \Diamond, \Diamond^\circ \rangle \geq -1$, and $\langle \Theta_F^{[k]}$, $\Theta_F^{\circ [2-k]}\rangle = -1$, it cannot be the case that both are vertically embedded at the same time. It hence remains to be shown that they also cannot both be non-vertically embedded at the same time. To see this, let us assume that both of them are non-vertically embedded. In this case, 
$\Theta^{[2-k]}$ is also a face of the polytope $\Delta$ constructed above \eqref{eq:defDelta}. However, as such it must have a dual face of dimension $k+1$ on $\Delta^\circ$ (defined in \eqref{eq:defDeltacirc}) which is furthermore vertical and contains $\Theta_F^{\circ [k]}$, violating our assumption that they are both non-vertically embedded. The only way out is that either $\Theta_F^{[2-k]}$ is vertically embedded (in which case it is not a face of $\Delta$) or that $\Theta_F^{\circ [k]}$ is vertically embedded and bounds the vertical face $\Theta^{\circ [k+1]}$ which then becomes the polar dual of 
$\Theta_F^{[2-k]}$ under the polar duality of $(\Delta,\Delta^\circ)$.

We are now in a position to discuss the structure of the normal fan $\Sigma_n(\Diamond)$. Consider the reflexive pair $\Delta,\Delta^\circ$, \eqref{eq:defDeltacirc} and \eqref{eq:defDelta}, and the normal fan of $\Delta$. Faces $\Theta_F^{\circ [k]}$ are dual to vertical faces $\Theta^{[3-k]}$ of $\Delta$ such that 
\begin{equation}
 \sigma_f(\Theta_F^{\circ [k]}) = \sigma_n(\Theta^{[3-k]}) \, .
\end{equation}
When we collapse $\Delta$ to $\Diamond$, $\Theta^{[3-k]}$ only stays a face if it extends below $F$, i.e. if the dual face $\Theta_F^{[2-k]}$ of $\Theta_F^{\circ [k]}$ under the polar duality of $(\Delta_F,\Delta_F^\circ)$ is vertically embedded. Correspondingly, the cone $\sigma_f(\Theta_F^{\circ [k]})$ is contained in the normal fan of $\Diamond$ if and only if $\Theta_F^{\circ [k]}$ is non-vertically embedded in $\Diamond^\circ$. 

If $\Theta_F^{\circ [k]}$ is vertically embedded in $\Diamond^\circ$, so that its dual $\Theta_F^{[2-k]}$ (under the polar duality of $(\Delta_F,\Delta_F^\circ)$) is non-vertically embedded, the cone $\sigma_f(\Theta_F^{\circ [k]})$ disappears when we collapse $\Delta$ to $\Diamond$. If we displace any face $\Theta_F^{[2-k]}$ of $\Delta_F$ by $m_0$ we find a face $\Theta_{F,m_0}^{[2-k]}$ of $\Delta$. If $\Theta_F^{[2-k]}$ is non-vertically embedded, 
$\Theta_{F,m_0}^{[2-k]}$ is identified with $\Theta_{F}^{[2-k]}$ when we collapse $\Delta$ to $\Diamond$ and correspondingly, $\sigma_n(\Theta_{F,m_0}^{[2-k]})$ is fused with $\sigma_n(\Theta_{F}^{[2-k]})$. These cones are the cones over the faces $\Theta_{F,\nu_0}^{\circ [k+1]}$ (which is the unique face connecting $\Theta_F^{\circ [k]}$ to $\nu_0$) and $\Theta_{F,+}^{\circ [k+1]}$ (which is the unique face extending $\Theta_F^{\circ [k]}$ vertically in $\Diamond^\circ$). 
\\
\\
We can hence summarize the discussion as follows: \\
\\
The normal fan of a top $\Diamond$ is equal to the face fan of $\Diamond^\circ$ except for vertically embedded faces $\Theta_F^{\circ [k]}$ and the faces $\Theta_{F,+}^{\circ [k+1]}$ and $\Theta_{F,\nu_0}^{\circ [k+1]}$ which are connected to them above and below $F$. For such faces, the normal fan $\Sigma_n(\Diamond)$ contains only a single $k+2$-dimensional cone which is the union of $\sigma_f(\Theta_{F,+}^{\circ [k+1]})$ and $\sigma_f(\Theta_{F,\nu_0}^{\circ [k+1]})$. Consequently, $\sigma_f(\Theta_F^{\circ [k]})$ is not present in $\Sigma_n(\Diamond)$ for vertically embedded faces $\Theta_F^{\circ [k]}$.
\\
\\
As we have described how to obtain the normal fan $\Sigma_n(\Diamond)$ by gluing cones in the face fan $\Sigma_f(\Diamond^\circ \cup \nu_0)$, it also follows directly that we can refine $\Sigma_n(\Diamond)$ to $\Sigma_f(\Diamond^\circ \cup \nu_0)$. 

As the fan $\Sigma$ is in turn a refinement of $\Sigma_f(\Diamond^\circ \cup \nu_0)$, we may think of the fan $\Sigma$
used to define the ambient space for $Z$ in the last section as a refinement of the normal fan of the lattice polytope $\Diamond$. Similar to \cite{1993alg.geom.10003B}, this allows to use the theory of \cite{DK} to study the geometry of $Z$. 

Note that \eqref{eq:conddualtop} implies that the support function defined by $\Diamond$, \eqref{eq:supfct}, is precisely such that $-[K_Z] = D_0$. Furthermore, any refinement of $\Sigma_n(\Diamond)$ which only introduces rays generated by lattice points on $\Diamond^\circ$ is crepant: the proper transform of the hypersurface equation is simply given by \eqref{eq:topmonos} again and the anticanonical class stays $D_0$. Refinements introducing higher-dimensional cones are trivially crepant.

\subsection{Topological Properties of Building Blocks}\label{sect:NandK}

Having discussed the detailed structure of the normal fan $\Sigma_n(\Diamond)$, we can now start to discuss the topological properties of the building blocks $Z$ we have conastructed. In order to find a matching and then compute the Betti numbers \eqref{eq:bettiG2} of the resulting $G_2$ manifold, we need to know the Hodge number $h^{2,1}(Z)$, as well as the lattices
\begin{equation}\label{eq:defNK}
 \begin{array}{ll}
 N &= im(\rho) \\
 K &= ker(\rho)/[S_0] 
 \end{array} \hspace{1cm} \rho: H^{1,1}(Z,\mathbb{Z}) \rightarrow H^{1,1}(S_0,\mathbb{Z})
\end{equation}
In this section we compute $N$, the rank of $K$, as well as the Hodge numbers of a building block $Z$ constructed from a pair of tops
$\Diamond, \Diamond^\circ$. As the building blocks discussed in this paper are constructed from a refinement $\phi: \Sigma \rightarrow \Sigma_n(\Diamond)$ of the normal fan of the polytope $\Diamond$, the theory of \cite{DK} can be applied.

\subsubsection{The Stratification Associated with a Refinement of the Normal Fan}

Every toric variety $\P_\Sigma$ enjoys a stratification into algebraic tori $(\C^*)^d$ which is determined by its fan $\Sigma$ \cite{danilov}.
For every $k$-dimensional cone $\sigma \in \Sigma(k)$ of the fan of an $n$-dimensional (complex) toric variety, there exists a corresponding 
$n-k$-dimensional stratum $S_\sigma \equiv (\C^*)^{n-k}$ and we may write
\begin{equation}
 \P_\Sigma = \amalg_{\sigma_k \in \Sigma} \,\,  (\C^*)^{n-k} \, .
\end{equation}
Furthermore a toric divisor $Y_i$, associated with a ray (1-dimensional cone) $\sigma_i$, has a stratification
\begin{equation}
 Y_i = \amalg_{\sigma_k \supseteq \sigma_i}  \,\,  (\C^*)^{n-k}  \, ,
\end{equation}
which is not unexpected as any toric divisor is again a toric variety (with fan $star(\sigma_i)$). 
A ray $\sigma_i$ corresponds to a homogeneous coordinate $z_i$ and, for simplicial fans, the closure of the associated stratum
$S_{\sigma_i}$ is simply given by setting $z_i = 0$. Similar results hold for
toric submanifolds of higher codimension.

If we consider an algebraic submanifold $Z \subset P_\Sigma$ of such a toric variety, we may intersect each stratum with $Z$ and thereby obtain
a stratification of $Z$:
\begin{equation}
 Z = \amalg_{\sigma_k \in \Sigma} \,\, S_\sigma \cap Z \equiv \amalg_{\sigma \in \Sigma} Z_\sigma\, .
\end{equation}
As each cone in the normal fan $\Sigma_n(\Diamond)$ is associated with a face of $\Diamond$, the (in general singular) hypersurface $Z_s$ in $\mathbb{P}_{\Sigma_n(\Diamond)}$ has a stratification associated with the face structure of $\Diamond$,
\begin{equation}\label{eq:stratnf}
Z_s = \amalg_{\Theta \in \Sigma_n(\Diamond)} Z_{\Theta} \, .
\end{equation}
Each $k$-dimensional face $\Theta^{[k]}$ of $\Diamond$ is associated with a cone $\sigma_n(\Theta^{[k]})$ of dimension $4-k$ in $\Sigma_n(\Diamond)$ and
$Z_{\Theta}$ is the $k-1$ dimensional (smooth) affine stratum $Z_s \cap (\C^*)^{k}$ corresponding to $\sigma_n(\Theta^{[k]})$. As we have discussed above, the fan $\Sigma$ giving us the ambient space $\P_\Sigma$ of our resolved hypersurface $Z$ is a refinement of $\Sigma_n(\Diamond)$. Here, the resolution $\phi_\mathbb{P}: \P_\Sigma \rightarrow \P_{\Sigma_n(\Diamond)}$ associated with a refinement $\phi: \Sigma \rightarrow \Sigma_n(\Diamond)$ induces a resolution $\phi_Z: Z \rightarrow Z_s$. This point of view has the advantage that we can describe the stratification of $Z$ as
\begin{equation}\label{eq:stratZ}
 Z = Z_{\Diamond} \amalg_{\Theta^{[3]} \in \Diamond} Z_{\Theta^{[3]}}  \amalg_{\Theta^{[2]} \in \Diamond}  E_{\Theta^{[2]}} \times Z_{\Theta^{[2]}}
  \amalg_{\Theta^{[1]} \in \Diamond}  E_{\Theta^{[1]}} \times Z_{\Theta^{[1]}} \, .
\end{equation}
Here $E_{\Theta^{[4-k]}}$ is the exceptional set originating from the refinement of the cone $\sigma_n(\Theta^{[4-k]})$
\begin{equation}
 E_{\Theta^{[4-k]}} = \amalg_{i=0}^{4-k-1} (\C^*)^{i} \, .
\end{equation}
For every $l$-dimensional cone of $\Sigma$ interior to $\sigma_n(\Theta^{[4-k]})$, there is a stratum $(\C^*)^{k-l}$ in $E_{\Theta^{4-k}}$. 

\subsubsection{Computing Hodge Numbers from Stratification}\label{sect:hodgefromstrata}

The reason for introducing all of this terminology and studying the normal fan in detail, is that the Hodge numbers of $Z$ can be computed from the Hodge numbers of its strata, which in turn have a combinatorial characterization \cite{DK}. The key is using the `Euler-characteristic' associated with
the mixed Hodge structure of cohomology with compact support:
\begin{equation}
e^{p,q}(X) \equiv \sum_k (-1)^k h^{p,q}(H^k_c(X)) \, .
\end{equation}
The numbers $e^{p,q}(X)$ become equal to the $(-1)^{p+q}h^{p,q}(X)$ for smooth compact varieties and enjoy the properties
\begin{align}
e^{p,q}(X_1 \amalg X_2) & = \, e^{p,q}(X_1) + e^{p,q}(X_2) \\
e^{p,q}(X_1 \times X_2) & = \sum_{\substack{p_1+p_2 = p \\ q_1+q_2=q}} e^{p_1,q_1}(X_1)\, \cdot \,e^{p_2,q_2}(X_2)
\end{align}
so that knowledge of the $e^{p,q}$ for all strata $Z_\sigma$ allows the computation of the Hodge numbers of $Z$.

Using the methods of \cite{DK} it is straightforward to compute the numbers 
\begin{equation}
 e^{p,q}(Z) \equiv \sum_k (-1)^k h^{p,q}(H^k_c(Z)) \, .
\end{equation}
as we discuss now.

The Hodge-Deligne numbers of the strata are determined by \cite{DK}
\begin{align}
 e^{p,0}(Z_{\Theta^{[k]}}) = (-1)^{k-1} \sum_{\Theta^{[p+1]}\leq \Theta^{[k]}} \ell^*(\Theta^{[p+1]})
\end{align}
for $p>0$ as well as
\begin{equation}\label{eq:DKsumrule}
 (-1)^{k-1} \sum_q e^{p,q}(Z_{\Theta^{[k]}}) = (-1)^p \binom{k}{p+1} + \varphi_{k-p}(\Theta^{[k]}) \, .
\end{equation}
Here, $\ell^*$ counts lattice points contained in the relative interior of a face and the functions $\varphi_n$ are defined as
\begin{equation}
\varphi_n(\Theta^{[k]}) := \sum_{j = 1}^n (-1)^{n+j} \binom{k+1}{n-j}\ell^*(j\Theta) \, ,
\label{eq:def-varphi}
\end{equation}
where $j \Theta$ stands for the polytope which is obtained by scaling all vertices of the face $\Theta$ 
by $j$ and then taking the convex hull. Note that simply
\begin{equation}
 \varphi_1(\Theta) = \ell^*(\Theta) \, .
\end{equation}

Let us introduce the following notation: we denote the number of points on the $n$-skeleton of $\Theta$ by
$\ell^n(\Theta)$. We can use this to rewrite e.g. the sum over the number of interior points of all $n$-dimensional
faces as $\ell^n(\Theta) - \ell^{n-1}(\Theta)$. Using this notation
\begin{equation}\label{eq:e00Z}
 e^{0,0}(Z_{\Theta^{[k]}}) = (-1)^{k-1}\left( \ell^1(\Theta^{[k]}) - 1 \right) \, .
\end{equation}

For a face of dimension $k \geq 4$ we also have that
\begin{equation}\label{eq:e31}
 e^{k-2,1}(Z_\Theta^{[k]}) = (-1)^{k-1}\left(\varphi_2(\Theta^{[k]}) -  \sum_{\Theta^{[k-1]} \leq \Theta^{[k]}} \varphi_1(\Theta^{[k-1]}) \right) .
\end{equation}
Finally high Hodge numbers, $ p+q\geq n$, satisfy
\begin{equation}
 e^{p,q}(Z_{\Theta^{[n]}}) =  \delta_{p,q}(-1)^{n+p+1}\binom{n}{p+1} \, .
\end{equation}

As a preparation for later sections, let us derive the Hodge-Deligne numbers $e^{p,q}$ of strata $Z_{\Theta^{[k]}}$ for $k \leq 4$.  As a corollary
of \eqref{eq:e00Z}, $e^{0,0}(Z_{\Theta^{[1]}}) = \ell^*(\Theta^{[1]})+1$, i.e. the stratum $Z_{\Theta^{[1]}}$ consists of $\ell^*(\Theta^{[1]})+1$ points.
Hence
\begin{equation}
e^{p,q}(Z_{\Theta^{[1]}}) =
\begin{array}{|cc}
  0 & 0 \\
 \ell^*(\Theta^{[1]})+1 & 0 \\
 \hline
\end{array}
\end{equation}
For $Z_{\Theta^{[2]}}$, we immediately find $e^{1,1}(Z_{\Theta^{[2]}}) =1 $. Furthermore, $e^{1,0}(Z_{\Theta^{[2]}}) = - \ell^*(\Theta^{[2]})$, and we can write
\begin{equation}
 e^{p,q}(Z_{\Theta^{[2]}}) = 
 \begin{array}{|cc}
 - \ell^*(\Theta^{[2]}) & 1 \\
1 - \ell^1(\Theta^{[2]}) & - \ell^*(\Theta^{[2]}) \\
 \hline
 \end{array}
\end{equation}

Similarly, we find for $Z_{\Theta^{[3]}}$ that
\begin{equation}
e^{p,q}(Z_{\Theta^{[3]}}) = \begin{array}{|ccc}
  \ell^*(\Theta^{[3]}) & 0 & 1 \\
\ell^2(\Theta^{[3]}) - \ell^1(\Theta^{[3]})  & -3 + \ell^*(2\Theta^{[3]})-4\ell^*({\Theta^{[3]}}) - \ell^2(\Theta^{[3]}) + \ell^1(\Theta^{[3]})& 0 \\
\ell^1(\Theta^{[3]}) - 1 & \ell^2(\Theta^{[3]}) - \ell^1(\Theta^{[3]})  &   \ell^*(\Theta^{[3]}) \\
\hline 
 \end{array}
\end{equation}

Finally, let us compute some of the Hodge-Deligne numbers of $Z_{\Theta^{[4]}}$. We have that
\begin{align}
 e^{0,0}(Z_{\Theta^{[4]}}) & = 1 - \ell^1(\Theta^{[4]}) \\
 e^{1,0}(Z_{\Theta^{[4]}}) & =  -\ell^2(\Theta^{[4]}) + \ell^1(\Theta^{[4]}) \\
 e^{2,0}(Z_{\Theta^{[4]}}) & = - \ell^3(\Theta^{[4]}) + \ell^2(\Theta^{[4]}) \\
 e^{2,1}(Z_{\Theta^{[4]}}) & = \ell^3(\Theta^{[4]}) - \ell^2(\Theta^{[4]}) - \varphi_2(\Theta^{[4]}) \\
 e^{2,2}(Z_{\Theta^{[4]}}) & = -4 
\end{align}

In general, the toric variety $\P_{\Sigma_n(\Diamond)}$ obtained from the normal fan $\Sigma_n(\Diamond)$, and correspondingly the hypersurface
$Z$ is not smooth and we need to resolve it. This can be done torically by refining the fan $\pi: \Sigma \rightarrow \Sigma_n(\Diamond)$. 
As we will always work with smooth hypersurfaces in practice, we continue to refer to the resolved hypersurface by $Z$.

The topology of $Z_\sigma = Z \cap S_\sigma$ solely depends the dimension of $\sigma$ and on the position of $\sigma$ within the normal fan $\Sigma_n(\Delta)$. 
If $\sigma$ is an $l$-dimensional cone contained in the relative interior of the $k$-dimensional cone of $\Sigma_n(\Diamond)$ associated with a face $\Theta^{[n-k]}$ we can write
\begin{equation}\label{eq:masterstratres}
 Z_\sigma = Z_{\Theta^{[n-k]}}\,\times \, (\C^*)^{k-l}\, . 
\end{equation}

In order to compute Hodge numbers we hence also need to know the Hodge-Deligne numbers of $(\C^*)^n$, they are simply given by
\begin{equation}
e^{p,q}((\C^*)^n) = \delta_{p,q} (-1)^{n+p} \binom{n}{p} \, . 
\end{equation}
For a point, i.e. $d=0$, we have that all $e^{p,q}(pt)$ vanish except $e^{0,0}(pt)=1$.

\subsubsection{Hodge Numbers of Building Blocks}

To compute the Hodge numbers of a building block $Z$, we start by noting that $Z$ enjoys a stratification
\begin{align}\label{eq:stratificationofZ}
 Z = Z_{\Diamond} \amalg Z_{\Theta^{[3]}} \amalg Z_{\Theta^{[2]}} \times [ \underbrace{\sum \C^*}_{\mbox{points in } \sigma_n(\Theta^{[2]})} 
 + \underbrace{\sum pt}_{\mbox{1-simplices on } \sigma_n(\Theta^{[2]})} ]\nn \\ \amalg Z_{\Theta^{[1]}} \times [ \underbrace{\sum (\C^*)^2}_{\mbox{points on }\sigma_n(\Theta^{[3]})}  + \underbrace{\sum \C^*}_{\mbox{1-simplices on }\sigma_n(\Theta^{[3]})} + \underbrace{\sum pt}_{\mbox{2-simplices on } \sigma_n(\Theta^{[3]})} ]
\end{align}
In the above expression, `points/n-simplices in $\sigma$' for some cone $\sigma$ is meant to be read as `points/n-simplices in the relative interior of $\sigma$ which are lattice points on $\Diamond^\circ \cup \nu_0$. The contribution of various strata of $\P_\Sigma$ crucially depends on the location of the associated points/n-simplices in the normal fan $\Sigma_n(\Diamond)$. As we have worked out the relation between faces of $\Diamond^\circ$ and cones in the normal fan $\Sigma_n(\Diamond)$ in Section \ref{sect:normalfanstops}, we can combine \eqref{eq:stratificationofZ} with the relations of Section \ref{sect:hodgefromstrata} to find the Hodge numbers of $Z$. 

To familiarize the reader with this method of computation, let us start with Hodge numbers $h^{i,0}(Z)$ which are relatively simple to compute. 
First note that only $Z_{\Diamond}$ potentially contributes to $e^{3,0}$ and $\ell^*(\Diamond) = 0$, so that we directly find $h^{3,0}(Z) = 0$. Next, we have
\begin{align}
e^{2,0}(Z) & = e^{2,0}(Z) + \sum_{\Theta^{[2]}<\Diamond} e^{2,0} (Z_{\Theta^{[2]}}) \nn \\
& = -\ell^3(\Diamond) + \ell^2(\Diamond) + \sum_{\Theta^{[2]}<\Diamond} \ell^*(\Theta^{[2]}) \nn \\
& = 0
\end{align}
and 
\begin{align}
 e^{1,0}(Z) & = e^{1,0}(Z) + \sum_{\Theta^{[3]}<\Diamond} e^{1,0} (Z_{\Theta^{[3]}}) + \sum_{\Theta^{[2]}<\Diamond} e^{1,0} (Z_{\Theta^{[2]}}) \nn \\
 & = -\ell^2(\Diamond) + \ell^1(\Diamond) + \sum_{\Theta^{[3]}<\Diamond} \left(\ell^2(\Theta^{[3]}) - \ell^1(\Theta^{[3]}) \right)
 - \sum_{\Theta^{[2]}<\Diamond} \ell^*(\Theta^{[2]}) \nn \\
 & = 0
\end{align}
In the last computation, we have used that each three-dimensional face of $\Diamond$ bounds precisely two two-dimensional faces, so that the sum over three-dimensional faces contributes $2(\ell^2(\Diamond)-\ell^1(\Diamond))$. 

Let us now move to the more involved parts of our computation. For $h^{1,1}(Z) = h^{2,2}(Z)$ we have
\begin{align}
 h^{1,1}(Z) & = e^{2,2}(Z_{\Diamond}) + \sum_{\Theta^{[3]}<\Diamond} e^{2,2}(\Theta^{[3]}) + \sum_{\Theta^{[2]}<\Diamond}e^{1,1}(\Theta^{[2]})\ell^*(\sigma_n(\Theta^{[2]})) \nn \\
& \hspace{1cm} + \sum_{\Theta^{[1]}<\Diamond}e^{0,0}(\Theta^{[1]})\,\cdot\, \ell^*(\sigma_n(\Theta^{[1]}))\, \cdot\,e^{2,2}((\C^*)^2) \nn \\
 & = -4 + \sum_{\Theta^{[3]}<\Diamond} 1 + \sum_{\Theta^{[2]}<\Diamond} \ell^*(\sigma_n(\Theta^{[2]})) + \sum_{\Theta^{[1]}<\Diamond}(\ell^*(\Theta^{[1]})+1)\ell^*(\sigma_n(\Theta^{[1]})) 
\end{align}
We use $\ell^*(\sigma)$ to denote the number of integral points in the relative interior of $\sigma$ which are also lattice points on $\Diamond^\circ \cup \nu_0$. 

This equation is of course similar to the equation for $h^{1,1}$ of a toric Calabi-Yau hypersurface \cite{1993alg.geom.10003B} (see also Appendix \ref{app:hodge_strat}) and it shares the same interpretation. Every lattice point on $\Diamond$ gives rise to a divisor on the ambient space $\P_\Sigma$, but only those which are not in the interior of a four-dimensional cone $\sigma_n(\Theta^{[0]})$ of the normal fan $\Sigma_n(\Diamond)$ also give a divisor on $Z$. In case a lattice point is in the relative interior of a three-dimensional cone $\sigma_n(\Theta^{[1]})$ of $\Sigma_n(\Diamond)$, it has $\ell^*(\Theta^{[1]}) +1$ irreducible components on $Z$ which give linearly independent divisors on $Z$. Finally, there are $4$ linear relations between the divisors on $Z$, which descend from the linear relations on $\P_\Sigma$, which we have to substract. 

Similarly, we can compute $h^{2,1}(Z) = -e^{2,1}(Z)$. As $e^{2,1}(Z_{\Theta^{[3]}}) = 0$, we find 
\begin{align}
e^{2,1}(Z) & =  e^{2,1}(Z_{\Diamond}) + \sum_{\Theta^{[2]} < \Diamond} e^{1,0}(Z_\Theta^{[2]})\,\cdot\, \ell^*(\sigma_n(\Theta^{[2]}))\,\cdot\,e^{1,1}(\C^*) \\
 & = \ell^3(\Diamond) - \ell^2(\Diamond) - \varphi_2(\Diamond) -  \sum_{\Theta^{[2]}<\Diamond} \ell^*(\Theta^{[2]})\ell^*(\sigma_n(\Theta^{[2]}))\\
 & = -\ell(\Diamond) + \ell(\Delta_F) - \sum_{\Theta^{[2]} < \Diamond} \ell^*(\Theta^{[2]})\cdot \ell^*(\sigma(\Theta^{[2]})) + \sum_{\Theta^{[3]} < \Diamond}\ell^*(\Theta^{[3]}) \\
 & = -h^{2,1}(Z)  \, .
\end{align}
Here we have used that 
\begin{equation}
 \varphi_2(\Diamond) = - 5 \ell^*(\Diamond) + \ell^*(2 \Diamond) = \ell(\Diamond) - \ell(\Delta_F) \, .
\end{equation}
for a projecting top $\Diamond$ and $\ell^3(\Diamond) - \ell^2(\Diamond)$ precisely counts the lattice points on $\Diamond$ in the relative interior of three-dimensional faces of $\Diamond$. 

One may also easily derive formulae for the Hodge numbers of toric divisors $Y_i$ on $Z$. We have done this for Calabi-Yau hypersurfaces in Appendix \ref{app:hodge_strat} and the formulae derived there can be straightforwardly used for toric divisors of building blocks $Z$
constructed from a pair of tops. The crucial input is the location of the point $\nu_i$ in the normal fan 
$\Sigma_n(\Diamond)$ and one only needs to replace `vertex' (`contained in 1D face', `contained in 2D face') by 
`contained in one-dimensional cone' (`contained in two-dimensional cone',`contained in three-dimensional cone') 
of $\Sigma_n(\Diamond)$ to make sense of the formulae. 

\subsubsection{The Lattices $N$ and $K$}

Let us now discuss the lattices $N$ and $K$. As already remarked in Section \ref{sect:adhocconstruction} we associate the embedding of the fibre $S$ with the divisor 
$D_0 = \{z_0 = 0\}$. A generic fibre is given by a $K3$ hypersurface $X_{(\Delta_F,\Delta^\circ_F)}$ defined via the reflexive pair $\Delta_F,\Delta^\circ_F$, see Appendix \ref{sect:mirk3}. For such a $K3$ surface, there are two interesting lattices: $\mbox{Pic}(X_{(\Delta_F,\Delta^\circ_F)})$ contains all divisors, whereas $\mbox{Pic}_{tor}(X_{(\Delta_F,\Delta^\circ_F)})$ only contains divisors which originate from divisors of the ambient space. Correspondingly, $\mbox{Pic}(X_{(\Delta_F,\Delta^\circ_F)})$ is contained in the Picard group of every fibre of $Z$ and 
$\mbox{Pic}_{tor}(X_{(\Delta_F,\Delta^\circ_F)})$ is generated by restricting all divisors corresponding to points on $\Delta^\circ_F$ to $X_{(\Delta_F,\Delta^\circ_F)}$. As every such point gives rise to a divisor on the threefold $Z$ as well, $\mbox{Pic}_{tor}(X_{(\Delta_F,\Delta^\circ_F)})$ must be contained in $N$. Note that there is a two-dimensional cone in $\Sigma$ generated by $\nu_i, \nu_0$ for every lattice point $\nu_i$ on $\Delta^\circ_F$. We may think of these two-dimensional cones as the divisors on $S$ in the image of $\rho$.

Things are slightly more complicated for other generators of $\mbox{Pic}(X_{(\Delta_F,\Delta^\circ_F)})$. Depending on the details of
$(\Diamond,\Diamond^\circ)$, these may or may not be contained in $N$. Whenever there is a one-dimensional face\footnote{We denote faces of $\Diamond$ or $\Diamond^\circ$ lying on $F$ by a subscript $F$. As we assume $\Diamond^\circ$ (and hence also $\Diamond$) to be projecting throughout this article, the set of all such fases is identical to the faces of $\Delta_F$ and $\Delta_F^\circ$, respectively.} $\Theta_F^{\circ [1]}$ on $\Delta^\circ_F$ such that, for $\Theta_F^{[1]}$ the dual face on $\Delta_F$, the product 
\begin{equation}\label{eq:picvspictor}
\ell^*(\Theta_F^{\circ[1]})\ell^*(\Theta_F^{[1]})\,  
\end{equation}
is non-zero, there is an extra contribution to $\mbox{Pic}$ which is not contained in $\mbox{Pic}_{tor}$. Let us consider a point $\nu_i$ on $\Delta^\circ_F$ contained in the interior of a face $\Theta^{\circ[1]}$ for which \eqref{eq:picvspictor} does not vanish. This signals that on $S$ we can write
\begin{equation}
 D_i  = \sum_\alpha D_i^\alpha \, ,
\end{equation}
and the $D_i^\alpha$ will be contained in $N$ if only if the above relation also holds on $Z$. Whereas the reducibility of toric divisors for 
$X_{(\Delta_F,\Delta^\circ_F)}$ corresponding to interior points of two-dimensional cones of $\Sigma_n(\Delta_F)$, comes from interior points of the dual one-dimensional face $\Theta_F^{\circ[1]}$ of $\Delta_F$, we have to consider the normal fan $\Sigma_n(\Diamond)$ in the case of $Z$. Here, only those toric divisors which correspond to interior points of three-dimensional cones of $\Sigma_n(\Diamond)$ can become reducible. For every one-dimensional face of $\Delta_F^\circ$, which forms a one-dimensional face $\Theta_F^{\circ[1]}$ of $\Diamond^\circ$, we must hence discriminate whether they sit in
two- or three-dimensional cones of the normal fan $\Sigma_n(\Diamond)$. As we have discussed in Section \ref{sect:vertnvert}, there is a simple rule: 
the cone over any face $\Theta_F^{\circ [i]}$ of $\Diamond^\circ$ laying on $F$ is contained in the normal fan of $\Diamond$ if and only if it is bounded (above F) by non-vertical faces of $\Diamond^\circ$. If $\Theta_F^{\circ [i]}$ is contained in a vertical face, then it is contained in a cone of dimension $i+2$ of the normal fan. This $i+2$-dimensional cone is the cone associated with its dual face $\Theta_F^{[1]}$ under the polar duality of $(\Delta_F,\Delta_F^\circ)$.  In this situation, the divisors associated with lattice points on $\Theta^{\circ [1]} = \Theta_F^{\circ [1]}$ have $\ell_F^*(\Theta^{[1]})+1$ irreducible components both in the generic $K3$ fibre and on the building block $Z$.
Hence
\begin{itemize}
 \item If $\Theta_F^{\circ[1]}$ bounds only non-vertical faces, only the contribution 
 to $\mbox{Pic}_{tor}$ associated with $\Theta_F^{\circ[1]}$ is in the image of $\rho$.
 \item If $\Theta_F^{\circ[1]}$ bounds a vertical face, the image of $\rho$ contains the lattice 
 \begin{equation}\label{eq:Llatfacet1}
L_i =  A_{\ell^*(\Theta_{F,i}^{\circ[1]})}^{\oplus \ell^*(\Theta_{F,i}^{[1]})}\, .
 \end{equation}
 besides the contribution to $\mbox{Pic}_{tor}$. 
\end{itemize}

%

Note that the lattice $N$ is always primitively embedded into $\mbox{Pic}(X_{(\Delta_F,\Delta^\circ_F)})$. This can be seen as follows: $N$ 
is given as $N = M^\perp$ in $\mbox{Pic}(X_{(\Delta_F,\Delta^\circ_F)})$, where $M$ is generated by all lattice elements if the form
\begin{equation}
 D_i^\alpha - D_i^\beta
\end{equation}
for all $\alpha,\beta$ and with $i$ running over all faces $\Theta_F^{\circ[1]}$ contained in a two-dimensional cone of $\Sigma_n(\Diamond)$. 
By construction, an orthogonal complement is primitively embedded. 

Using the label (vb) for vertically bounded faces $\Theta_{F,i}^{\circ[1]}$, the lattice $N$ is hence 
\begin{equation}
 N = \mbox{Pic}_{tor}(X_{\Delta_F,\Delta^\circ_F}) + \sum_{vb \,\, \Theta_{F,i}^{\circ[1]}} L_i
\end{equation}
and its rank is
\begin{equation}
rk(N) = |\mbox{Pic}_{tor}(X_{\Delta_F,\Delta^\circ_F})| + \sum_{vb \,\,\Theta_{F}^{\circ[1]} }  \ell^*(\Theta_F^{\circ [1]})\ell^*(\Theta_F^{[1]}) \, .
\end{equation}
with 
\begin{equation}
|\mbox{Pic}_{tor}(X_{\Delta_F,\Delta^\circ_F})| = \ell^1(\Delta^\circ_F) - 3  \, .
\end{equation}
The rank of $K$ can now be found from
\begin{equation}
\mbox{rk}(K) =h^{1,1}(Z)-\mbox{rk}(N)-1 \, . 
\end{equation}
The divisors contributing to $K$ correspond to singular fibre components, which in turn correspond to lattice points on $\Diamond$ above $F$
(see \eqref{eq:fibrecomp}) as well as points interior to two-dimensional faces of $\Delta^\circ_F$.


\subsection{Summary}\label{sect:sum_proof}

Let us summarize the main points of the construction presented above and finish the proof that
it indeed leads to a building block as defined by \cite{corti_g2}, i.e. show that it satisfies the requirements i) to iv) in \eqref{sect:defsbuildbocks}.

Starting from a projecting top $\Diamond^\circ$ with
$\Diamond^\circ\cap F =\Delta^\circ_F$, $F = m_0^\perp$, there is a dual\footnote{Here, our notation $^\circ$
is meant to indicate `dual' rather than `polar dual'.} top $\Diamond$:
\begin{equation}\label{eq:topsduality}
 \begin{aligned}
 & \langle \Diamond, \Diamond^\circ \rangle \geq -1 & \\
  \langle  \Diamond,\nu_0 \rangle \geq 0 \hspace{.5cm} & &  \langle m_0, \Diamond^\circ\rangle \geq 0
\end{aligned}
\end{equation}
and we may choose coordinates such that $m_0 = (0,0,0,1)$ and $\nu_0 = (0,0,0,-1)$.
As a convex lattice polytope, $\Diamond$ defines a toric variety $\P_{\Sigma_n(\Diamond)}$ via a normal fan $\Sigma_n(\Diamond)$, as well as a line bundle $\mathcal{O}(D_\Diamond)$ on it. The face fan $\Sigma(\Diamond\cup\nu_0)$ of $\Diamond^\circ \cup \nu_0$ is a refinement of $\Sigma_n(\Diamond)$. 

In general, $\P_{\Sigma_f(\Diamond\cup\nu_0)}$ will have singularities which meet $Z$. We may however, further refine the fan $\Sigma$ according to a appropriate\footnote{We would like such triangulations to involve all lattice points on $\Diamond^\circ$ and give rise to a projective toric variety $\P_\Sigma$.)} triangulation of the $\Diamond^\circ$ to find a maximally crepant desingularisation. In our case of interest, where $Z$ is a threefold and $\P_\Sigma$ a fourfold, such a triangulation will only leave point-like singularities\footnote{The reason for this is that any fine triangulation of a face of dimension less than three leads to simplices of lattice volume unity.} in $\P_\Sigma$ which do not meet a generic hypersurface. The hypersurface $Z$ is then given by a generic section of $\mathcal{O}(\Diamond)$:
\begin{equation}\label{eq:zdefeq}
Z: \,\,\, 0 = \sum_{m \in \Diamond} z_0^{\langle m, \nu_0 \rangle}\prod_{\nu_i} z_i^{\langle m, \nu_i\rangle +1} \, .
\end{equation}
Its Hodge numbers are
\begin{align}\label{eq:hodgenumbersZ}
h^{1,1}(Z) & =  -4 + \sum_{\Theta^{[3]}<\Diamond} 1 + \sum_{\Theta^{[2]}<\Diamond} \ell^*(\sigma_n(\Theta^{[2]})) + \sum_{\Theta^{[1]}<\Diamond}(\ell^*(\Theta^{[1]})+1)\ell^*(\sigma_n(\Theta^{[1]}))  \nn \\
h^{2,1}(Z) & = \ell(\Diamond) - \ell(\Delta_F) + \sum_{\Theta^{[2]} < \Diamond} \ell^*(\Theta^{[2]})\cdot \ell^*(\sigma(\Theta^{[2]})) - \sum_{\Theta^{[3]} < \Diamond}\ell^*(\Theta^{[3]})\, ,
\end{align}
and the ranks of the lattices $N$ and $K$ defined in \eqref{eq:defNK} are
\begin{align}
 |N| &=  \ell^1(\Delta^\circ_F) - 3 + \sum_{vb \,\,\Theta_{F}^{\circ[1]} }  \ell^*(\Theta_F^{\circ [1]})\ell^*(\Theta_F^{[1]})  \nn \\
 |K| &= h^{1,1}(Z)-\mbox{rk}(N)-1 \, .
\end{align}

For a projecting top, $\Delta^\circ_F = \Diamond^\circ \cap F$ and 
$\Delta_F = \Diamond \cap F$ are a reflexive pair \cite{Avram:1996pj}. The hypersurface $Z$ 
is hence fibred by a $K3$ surface from the algebraic family defined by the reflexive pair $(\Delta^\circ_F, \Delta_F)$. 

We can now start with the proof of requirements i) to iv) in \eqref{sect:defsbuildbocks}
needed for a building block. 

\begin{itemize}
 \item[i)] This is true by construction. In particular, setting $z_0 = 0$, we 
find a K3 surface $S$ given as a hypersurface in a toric variety constructed from the reflexive pair $(\Delta_F , \Delta_F^\circ)$. Furthermore, we have that 
 \be
 [S] = \sum_{\nu_i \in \Diamond^\circ} \langle m_0, \nu_i \rangle D_i = D_0
 \ee
 which is equal to the class of the fibre of the fibration implicit in the top.
  \item[ii)] Denoting the lattice points on $\Diamond^\circ$ by $\nu_i$, the anticanonical class of $\P_\Sigma$ is $D_0 + \sum_{i} D_i$ and the class of $Z$ is $[Z] =  \sum_{i} D_i$. By adjunction we hence have that $-[K_Z] = D_0 = [S_0]$. As shown in Section \ref{sect:adhocconstruction}, the class $[S]$ is furthermore primitive in $H^{2}(Z,\mathbb{Z})$.
 \item[iii)] As we have remarked above, the lattice $N$ is primitively embedded in the Picard lattice of the generic fibre, $\mbox{Pic}(X_{(\Delta_F,\Delta^\circ_F)})$. As the Picard lattice is primitively embedded in $H^2(S,\Z)$, it follows that also $N$ is primitively embedded in $ H^2(S,\Z)$.
\item[iv)] 
As $Z$ is a projective algebraic manifold which is defined as a hypersurface by a Newton polytope $\Diamond$ and embedded in a toric variety $P_\Sigma$, where $\Sigma$ is a refinement of the normal fan $\Sigma(\Diamond)$, we may use the results of \cite{KrBat} and compute $\mbox{Tors}(H^3(Z,\mathbb{Z}))$ combinatorially as
\begin{equation}\label{eq:brauergroup}
\mbox{Tors}(H^3(Z,\mathbb{Z})) \cong \mbox{Hom}(\,\Lambda^2 {\bf N}/({\bf N} \wedge {\bf N}_\Diamond^{(2)})\, ,\, \mathbb{Q}/\mathbb{Z}\,).
\end{equation}
Here, ${\bf N}_{\Diamond}^{(2)}$ is generated by all lattice vectors $v$ in ${\bf N}$ such that the function 
$f_v \equiv \langle v, \ast \rangle$ on $\Diamond$ has a minimum along a face of dimension greater than two. 

Note that $f_{\nu_0}$ attains a minimum along the face $\Theta^{[3]}_0 = \Diamond \cap F$, so $\nu_0$ is one of the generators of $\N_\Diamond^{(2)}$. The quotient \eqref{eq:brauergroup} can hence only receive a non-trivial contribution from lattice vectors parallel to $F$. 
To show that \eqref{eq:brauergroup} vanishes, we make use of the following three key facts \cite{Avram:1996pj,Candelas:2012uu}:
\begin{itemize}
\item[a)] For a projecting top $\Diamond^\circ$, its dual $\Diamond$, \eqref{eq:topsduality}, is also projecting (see Section \eqref{sect:adhocconstruction}).
\item[b)] Any two projecting tops with the same $\Delta_F$ can be assembled into a reflexive polytope.
\item[c)] There is a known list of 16 reflexive polyhedra for which \eqref{eq:brauergroup} is non-trivial \cite{KrBat}. 
\end{itemize}
For any top $\Diamond$ we may use a) and b) to form a reflexive polytope $\Delta(\Diamond)$ using $\Diamond$ and a copy $\Diamond'$ with the fourth coordinate inverted. If the quotient \eqref{eq:brauergroup} is non-trivial for such a top $\Diamond$, it must also be non-trivial for $\Delta(\Diamond)$. 
However, out of all the four-dimensional reflexive polytopes, there are only sixteen cases for which \eqref{eq:brauergroup} is nontrivial \cite{KrBat} (see also \cite{He:2013ofa}). We have recorded this list in Appendix \ref{sect:polybrauer}. It is then possible to directly check if any of the sixteen relevant four-dimensional polytopes admits a sub-polytope cutting it into a pair of projecting tops. We have done this computation using SAGE \cite{sage}. Even though all but three of these polytopes do have reflexive subpolytopes, none of these turns out to be projecting. We have hence shown that \eqref{eq:brauergroup} must be trivial for any projecting top.

\end{itemize}

%
%
%

This completes the proof that our construction for $Z$ leads to a building block as defined in \cite{corti_g2} for any projecting top.

\subsection{Relation to Construction via Semi-Fano Threefolds}

In \cite{corti_weakfano}, building blocks for $G_2$ manifolds are constructed by appropriate blowups of so-called semi-Fano threefolds $A$, which
are a subclass of weak Fano threefolds.

A \emph{weak Fano threefold} is a non-singular projective complex threefold $A$ such that the anticanonical bundle $-K_A$ is big and nef. This means that
$-K_A \cdot C > 0$ for any curve $C$ in $A$ and $(-K_X)^3 = 2g-2>0$. The integer $g$ is called the genus of $A$.

For a weak Fano threefold $A$, there is an anticanonical morphism $\varphi: A \rightarrow A_{ac}$ to the anticanonical model $A_{ac}$ of $A$. If this map only contracts divisors to curves and curves to points (but not divisors to points) it is called \emph{semi-small} and, following \cite{corti_weakfano}, the threefolds for which this is the case are called \emph{semi-Fano threefolds}\footnote{This terms is also used with a different meaning in the mathematics literature.}.

For such threefolds, \cite{corti_weakfano} construct building blocks as follows. First, choose two anticanonical hypersurfaces $S$ and $S'$ meeting transversely. Then blow up $A$ along $S \cap S'$. In practice, we may accomplish this as follows. Let $S_0$ be defined by an equation $P_0=0$ and $S_\infty$ by $P_\infty=0$ in
$A$. The building block $Z$ is then described in $A\times \P^1$
\begin{equation}\label{eq:ccpblowup}
\xi_\infty P_0 = \xi_0 P_\infty \, ,
\end{equation}
where $[\xi_0:\xi_\infty]$ are homogeneous coordinates on the $\P^1$. 

In case $A$ is the toric variety associated to a reflexive polytope $\Delta^\circ_F$, such a manifold $Z$ is obtained from our construction by using the trivial top, i.e. the convex hull $\Diamond^\circ$ of $\Delta^\circ_F \cup (0,0,0,1)$. $A$ plays the role of the ambient (weak Fano) toric manifold in which the fibre $S_0$ is embedded, i.e. the (resolution of) the toric variety $\mathbb{P}_{\Delta_F}$. A toric threefold constructed from a three-dimensional reflexive polytope $\Delta^\circ$ is semi Fano if and only if $\Delta^\circ$ has no interior points to facets \cite{corti_weakfano}. Of the 4319 reflexive polytopes in three dimensions, $899$ have this property.

For semi-Fano threefolds $A$, the building blocks obtained as \eqref{eq:ccpblowup} have a trivial $K$ if the $K3$ surfaces $S_0$ and $S_\infty$ are smooth. Of course, one may choose e.g. a singular $S$ and then construct a crepant resolution of \eqref{eq:ccpblowup}, as done in \cite{Kovalev_Lee,corti_weakfano,Halverson:2014tya}. The strength of our construction is a systematic and convenient characterization of such degenerations, i.e. a framework to systematically enhance $K$ and hence $b_2$ of the resulting $G_2$ manifolds. Furthermore, a description in terms of polytopes lends itself to a straightforward description of singular transitions (at least on the level of the building blocks $Z$).

For the blowup \eqref{eq:ccpblowup} to give rise to a building block, it is sufficient to demand that $A$ is a weak Fano threefold. The relevance of the extra condition of being semi-Fano lies in the deformation theory of the anticanonical hypersurfaces of $A$. For any weak Fano threefold, an anticanonical hypersurface is a member of a family of lattice polarized $K3$ surfaces with polarizing lattice $L_A$. One may hence wonder whether the moduli space\footnote{Strictly speaking theses are moduli stacks.} $\mathfrak{F}^{A}$ of the embedded $K3$ surface $X \hookrightarrow A$ is as large as the moduli space of the lattice polarized family $\mathfrak{K}^{N_A}$. It turns out that this is indeed the case if $A$ is semi-Fano. In this case the forgetful morphism
\begin{equation}\label{eq:modhypvslpk3}
s:  \mathfrak{F}^{A} \rightarrow \mathfrak{K}^{L_A}
\end{equation}
is generically surjective \cite{corti_weakfano}.

If we are given two building blocks and have found primitive lattice embeddings which facilitate a glueing to a $G_2$ manifold, we furthermore have to be able to choose the holomorphic two-forms and K\"ahler forms such that they satisfy \eqref{eq:hkrot}. It is not hard to see if this can be satisfied for a pair of families of lattice polarized $K3$ surfaces, but we need that the $K3$ fibres of $Z_+$ and $Z_-$ can be choose appropriately. As these fibre are realized as hypersurfaces \eqref{eq:modhypvslpk3} is the crucial result allowing us to decide whether we can form the twisted connected sum.

\section{Examples}\label{sect:examples}

In this section, we discuss a few examples of building blocks obtained from tops in ascending order of complexity. 

In compactifications of M-Theory on manifolds of $G_2$ holonomy, the number of massless $U(1)$ gauge symmetries is determined by
the Betti number $b_2$. From \eqref{eq:bettiG2}, we may distinguish two different sources: $K_\pm$ and $ N_+ \cap N_-$. 

The dual cycles of the forms originating from the first part, $K_\pm$, are components of degenerate $K3$ fibres (times the extra $S^1$) and we can think of them as `localized' on $Z$. In particular, they do not play any role in the gluing to a $G_2$ manifold. Collapsing such fibre components (if possible) gives rise to $U(1)$ charged massless matter (as observed in \cite{Halverson:2014tya}). 

For the second part, $ N_+ \cap N_-$, the dual 5-cycles can be thought of as two-cycles fibred over the whole base $S^3$. 
Collapsing the fibres of such five-cycles (if possible) would give us singularities of codimension four, which give rise to gauge symmetries in compactifications of M-Theory. We can hence morally think of the $U(1)$s related to such cycles as the Cartan $U(1)$s of some non-abelian gauge group \cite{Halverson:2015vta}.

\subsection{Building Blocks with a Quartic K3 Fibre}\label{sect:quarticfibre}

Some of the simplest examples of building blocks use a quartic $K3$ surface embedded in $\P^3$ as the fibre. The trivial top is realized as
a polyhedron with the vertices
\begin{equation}\Diamond^\circ \sim
\left(\begin{array}{rrrrr}
-1 & 0 & 0 & 0 & 1 \\
-1 & 0 & 0 & 1 & 0 \\
-1 & 0 & 1 & 0 & 0 \\
0 & 1 & 0 & 0 & 0
\end{array}\right)
\end{equation}
In this case, the dual top has vertices 
\begin{equation}\Diamond \sim
\left(\begin{array}{rrrrrrrr}
-1 & -1 & 3 & 3 & -1 & -1 & -1 & -1 \\
-1 & -1 & -1 & -1 & 3 & 3 & -1 & -1 \\
-1 & -1 & -1 & -1 & -1 & -1 & 3 & 3 \\
-1 & 0 & 0 & -1 & 0 & -1 & 0 & -1
\end{array}\right) 
\end{equation}
and our construction simply gives a building block realized as a hypersurface of degree $(4,1)$ in $\P^3 \times \P^1$. 
There are $128$ degenerate and no reducible fibres. We can use the Lefschetz hyperplane theorem to find $h^{1,1}=2$ and $h^{1,0}=h^{2,0}=0$. Using adjunction we can also compute characteristic classes of $Z$, which then give us
\begin{equation}
 h^{1,1}(Z) = 2 \hspace{1cm} h^{2,1}(Z) = 33 \hspace{1cm} h^{3,0}(Z) = 0 \, .
\end{equation}
from the Euler characteristic and the arithmetic genus. The same values can be obtained from \eqref{eq:hodgenumbersZ}. As we have a quartic K3 surfaces as the fibre, we find $rk(N) = 1$ and hence $rk(K) = 0$, so that this building block does not give rise to any $U(1)$ of the `localized' type. 
\\
\\
\noindent
{\bf A degenerate fibre with four components}

We may make this example slightly more interesting by forcing a degenerate fibre given by
\begin{equation}
z_1 z_2 z_3 z_4 = 0 \, .
\end{equation}
This example has appeared in \cite{Kovalev_Lee,corti_weakfano,corti_g2,Halverson:2014tya} and we may reconstruct it starting from a pair of dual tops
with vertices
\begin{equation}\label{eq:quartic4compZ}
\Diamond^\circ_{1234} \sim \left(\begin{array}{rrrrrrrr}
-1 & -1 & 0 & 0 & 0 & 0 & 1 & 1 \\
-1 & -1 & 0 & 0 & 1 & 1 & 0 & 0 \\
-1 & -1 & 1 & 1 & 0 & 0 & 0 & 0 \\
0 & 1 & 0 & 1 & 0 & 1 & 0 & 1 \\
\nu_1 & \hat{\nu}_1 & \nu_2 & \hat{\nu}_2 & \nu_3 & \hat{\nu}_3 & \nu_4 & \hat{\nu}_4
\end{array}\right)\, , \hspace{1cm}
\Diamond_{1234} \sim \left(\begin{array}{rrrrr}
-1 & 3 & 0 & -1 & -1 \\
-1 & -1 & 0 & 3 & -1 \\
-1 & -1 & 0 & -1 & 3 \\
0 & 0 & -1 & 0 & 0 \\
\mu_1 & \mu_2 & \hat{\mu} & \mu_3 & \mu_4
\end{array}\right)\, .
\end{equation}
Here, we have implemented the fibre embedding by using the ${\bf N}$-lattice polytope corresponding to $\P^3$ as $\Delta^\circ_F$, and engineered four fibre components by placing four points on top of it. We have also indicated a labelling for the vertices of $\Diamond^\circ$.

Evaluating \eqref{eq:hodgenumbersZ} one finds that 
\begin{equation}
 h^{1,1}(Z) = 5 \hspace{1cm} h^{2,1}(Z) = 12\, .
\end{equation}
As the rank of $N$ remains $1$, we find $rk(K) = 3$. Hence a $G_2$ manifold which is constructed from such a building block as a twisted connected sum will receive three classes in $H^2(X)$ irrespective of the details of the gluing which in turn contribute three $U(1)$ factors to a compactification of M-Theory.

The classes in $K$ are contributed from the points $\hat{\nu}_i$ which correspond to the components of the reducible $K3$ fibre. The four divisors
$[\hat{z}_i]$ are rational surfaces\footnote{Their non-trivial Hodge numbers are $h^{1,1} = 1,5,9,13$.} meeting along six rational 
curves. The dual graph of these intersections is topologically $S^2$, so this is a fibre of type III in the classification of \cite{Kulikov,FM}.
Note that the lattice point $(0,0,0,1)$ on $\Diamond^\circ$ sits inside a four-dimensional cone of $\Sigma_n(\Diamond)$, so that it does not give rise to a divisor on $Z$.

Let us repeat some of the analysis of \cite{Halverson:2014tya} in the language of this paper.
The normal fan $\Sigma_n(\Diamond)$ is determined by its cones of maximal dimension, they are
\begin{equation}
 \langle \hat{\nu}_i, \hat{\nu}_j, \hat{\nu}_k, \nu_0 \rangle\, , \hspace{1cm} \forall i \neq j \neq k 
\end{equation}
as well as the cone $\langle \hat{\nu}_1, \hat{\nu}_2, \hat{\nu}_3, \hat{\nu}_4 \rangle$. Note that the coordinates $z_i$ related to the vertices 
$\nu_i$ of $\Diamond^\circ$ appear as exceptional divisors upon refinement of the normal fan $\Sigma_n(\Diamond)$ to the fan $\Sigma$. There are four curves $C_i$ which are given by $[\hat{z}_i] \cdot [z_0]$ on $Z$ before resolution (refinement of the normal fan). They correspond to two-dimensional cones of $\Sigma_n(\Diamond)$ spanned by $\hat{\nu}_i$ and $\nu_0$. As the dual two-dimensional faces on $\Diamond$ are the ones spanned by vertices $\mu_l$, $\mu_m$ and $\mu_n$, they have three interior points, so that the genus of the $C_i$ is three. Any two such curves meet in four points. In the fan refinement the  $\nu_i$ are introduced inside of the two-dimensional cones spanned by $\hat{\nu}_i$ and $\nu_0$. Hence we may think of the $[z_i]$ as being a fibration of $\P^1$ over the curves $C_i$, which have genus $3$. 

Note that we may also think of $Z$ as the blowup of a singular family in $\P^3 \times \P^1$. Thinking in this way, the coordinates $\hat{z}_i$ correspond to the exceptional divisors.

The $24$ rigid rational curves found in this geometry in \cite{Halverson:2014tya} can be seen as follows. There is a triangulation of $\Diamond^\circ$ such that for any pair of divisors $[z_i],[\hat{z}_j]$ with $i>j$, $[z_i]\cdot [\hat{z}_j]$ defines a curve $C_{ij}$ on $Z$. This triangulation corresponds to introducing the rays refining $\Sigma_n(\Diamond)$ to $\Sigma$ in the order $\nu_1$, $\nu_2$, $\nu_3$, and $\nu_4$.
The corresponding two-dimensional cones of $\Sigma$ sit inside three-dimensional cones of $\Sigma_n(\Diamond)$, which are in turn dual to one-dimensional faces $\Theta^{[1]}$ of $\Diamond$ with three interior points each. This means that
$C_{ij}$ has four irreducible components, each of which is a $\P^1$. We hence find $4 \cdot 6 = 24 $ such curves. Note in particular that these curves do not intersect the fibre $S_0$ as they are away from $\nu_0$. This means that we may think of them as being away from the region of $Z$ which is used in the gluing process to a $G_2$ manifold. 

The intersection numbers of these curves with the divisors $[\hat{z}_i]$ are
\begin{equation}
 \begin{array}{c|cccc}
  & [\hat{z}_1]  & [\hat{z}_2]  & [\hat{z}_3]  & [\hat{z}_4] \\
  \hline
 C_{12} & 1 & -1 & 0 & 0 \\
 C_{13} & 1 & 0 & -1 & 0 \\
 C_{14} & 1 & 0 & 0 & -1 \\
 C_{23} & 0 & 1 & -1 & 0 \\
 C_{24} & 0 & 1 & 0 & -1 \\
 C_{34} & 0 & 0 & 1 & -1
 \end{array}
\end{equation}
Each one of the divisors $[\hat{z}_i]$ defines a class in $H^{1,1}(Z,\Z)$ which is in the kernel $K$ of the restriction map to $S_0$ at $z_0=0$. If we glue the building block (times $S^1$) with an appropriate second building block, each of these divisors hence defines a class in $H^2(X,\Z)$ of the resulting $G_2$ manifold. For M-Theory compactifications on $X$, we can hence associate a $U(1)$ generator $Q_i$ with each of the $[\hat{z}_i]$. There is a linear relation which says that
\begin{equation}
 \sum_i [\hat{z}_i] = [z_0] = [S_0] \, .
\end{equation}
Hence these divisors only define three independent classes in $K$ and correspondingly the $Q_k$ only generate three independent $U(1)$s. States corresponding to wrapped M2 branes on the curves $C_{ij}$ are charged under those $U(1)$s with charges given by 
\begin{equation}
 \int_{C_{ij}} Q_k \, .
\end{equation}
These charges can be read off from the table above. A choice of independent $U(1)$ generators reproducing the charge matrix of \cite{Halverson:2014tya} is given by 
\begin{equation}
 [\hat{z}_1] - [\hat{z}_i] \, , \mbox{for}\,\, i = 2,3,4 \, .
\end{equation}

\subsection{Building Blocks with a K3 Fibre of Degree 2}\label{sect:sexticfibre}

Let us consider another simple type of $K3$ surface with $rk(N)=1$ as the fibre. We can realize a $K3$ surface $S$ with $Pic(S) = (2)$ 
as a double cover over $\P^2$. Constructing the corresponding polytope $\Delta^\circ_F$ as well as the trivial top and its dual is straightforward:
\begin{equation}
\Diamond^\circ\sim \left(\begin{array}{rrrrr}
-1 & 0 & 0 & 0 & 3 \\
0 & -1 & 0 & 0 & 1 \\
0 & 0 & -1 & 0 & 1 \\
0 & 0 & 0 & 1 & 0
\end{array}\right)\, , \hspace{1cm}
\Diamond\sim 
\left(\begin{array}{rrrrrrrr}
-1 & -1 & 1 & 1 & 1 & 1 & 1 & 1 \\
1 & 1 & 1 & 1 & 1 & 1 & -5 & -5 \\
1 & 1 & 1 & 1 & -5 & -5 & 1 & 1 \\
-1 & 0 & 0 & -1 & 0 & -1 & 0 & -1
\end{array}\right)
\end{equation}
As in the first example in Section \ref{sect:quarticfibre} above, choosing the trivial top in the ${\bf N}$ lattice gives a dual top with the maximal number of vertical faces. This example has an extra twist, though. The polytope $\Delta^\circ_F$ has an integral point interior to a 2-dimensional face, $(1,0,0,0)$. In the normal fan $\Sigma_n(\Diamond)$, this point sits inside a three-dimensional cone, so that there is a corresponding divisor on $Z$ but not on $S_0$. Hence we find $K \supset \mathbb{Z}$. The Hodge numbers of this example are
\begin{equation}
 h^{1,1}(Z) = 3 \hspace{1cm} h^{2,1}(Z) = 37\, .
\end{equation}

Note that the non-trivial $K$ can be understood in terms of a reducible fibre also in this case. We can describe the three-fold $Z$ by an equation of the form
\begin{equation}
(\xi_1 - \xi_0) z_1^2 + \hat{x} z_1 P_{1,3}(\xi,\xi_0,z_2,z_3,z_4) +  \hat{x}^4 P_{1,6}(\xi,\xi_0,z_2,z_3,z_4) = 0 \, .
\end{equation}
Here, $[\xi:\xi_0]$ are the coordinates of the base $\P^1$ corresponding to the vertices $(0,0,0,1)$ and $(0,0,0,-1)$. The coordinates
$[z_1:z_2:z_3:z_4]$ are the homogeneous coordinate of the weighted $\P^3_{1113}$ with weights $3,1,1,1$ and $[\hat{x}]$ is the exceptional divisor
of the blowup at $z_2=z_3=z_4=0$, corresponding to the ray over $(1,0,0,0)$. The polynomials $P$ have the indicated degrees under the $\C^*$ actions 
of the base $\P^1$ and the $\P^3_{1113}$. It follows that the fibre over $\xi_1 -  \xi_0=0$ is reducible and consists of two components.

We can construct a closely related model $Z'$ for which $K = 0$. Replacing the vertex $(0,0,0,1)$ with $(1,0,0,1)$ creates a vertical face on $\Diamond^{' \circ}$ which contains $(1,0,0,0)$ in its boundary. Correspondingly, the 2-dimensional face of $\Delta^\circ_F$ containing $(1,0,0,0)$ does not give rise to a cone in the normal fan of $\Diamond^{'}$. This means that now $(1,0,0,0)$ sits inside a 4-dimensional cone of $\Sigma_n(\Diamond^{'})$, so that there is no corresponding divisor on $Z'$. This is confirmed by computing the Hodge numbers from \eqref{eq:hodgenumbersZ}:
\begin{equation}
 h^{1,1}(Z') = 2 \hspace{1cm} h^{2,1}(Z') = 54\, .
\end{equation}

\subsection{An Example with \texorpdfstring{$Pic_{tor}(S) \neq Pic(S)$}{Lg}.}\label{sect:NneqPic}

Let us consider an example where there is a non-trivial correction term to the Picard lattice of the generic fibre. A simple such example is given by embedding the $K3$ fibre $S$ in the weighted projective space $\P^3_{1,1,2,2}$. The trivial top and its dual then have the vertices
\begin{equation}
\Diamond^\circ \sim \left(\begin{array}{rrrrr}
-1 & 0 & 0 & 0 & 2 \\
0 & -1 & 0 & 0 & 2 \\
0 & 0 & -1 & 0 & 1 \\
0 & 0 & 0 & 1 & 0
\end{array}\right)\, ,\hspace{1cm}
\Diamond \sim
\left(\begin{array}{rrrrrrrr}
-2 & -2 & 1 & 1 & 1 & 1 & 1 & 1 \\
1 & 1 & 1 & 1 & 1 & 1 & -2 & -2 \\
1 & 1 & 1 & 1 & -5 & -5 & 1 & 1 \\
-1 & 0 & 0 & -1 & 0 & -1 & 0 & -1
\end{array}\right)
\end{equation}
The hodge numbers of $Z$ are
\begin{equation}
 h^{1,1}(Z) = 3 \hspace{1cm} h^{2,1}(Z) = 28\, .
\end{equation}

Let us first discuss the geometry of the fibre $S$ and the polytope $\Delta^\circ_F$ in some detail. $\Delta^\circ_F$ has four vertices
and the number of interior points of the dual two-dimensional faces is 
\begin{equation}
 \begin{array}{r|r|r}
 \mbox{vertex} & \mbox{Divisor} & \ell^*(\Theta^{[2]}) \\
 \hline
(2,2,1,0) & D_1 & 1 \\
(0,0,-1,0)& D_2 & 1 \\
(-1,0,0,0)& D_3 & 4 \\
(0,-1,0,0)& D_4 & 4
 \end{array}\, .
\end{equation}
Hence the divisors corresponding to the first two vertices are elliptic curves and the divisors corresponding to the last two vertices have genus
4. Their self-intersection is $D_1 \cdot D_1 = D_2 \cdot D_2 = 0$ and $D_3 \cdot D_3 = D_4 \cdot D_4= 6$. The polytope $\Delta^\circ_F$ has a further integral point at $(1, 1, 0,0)$ which sits in between $\nu_1$ and $\nu_2$. The dual of this edge on $\Delta_F$ has vertices $ (-2, 1, 1,0)$ and  $(1, -2, 1,0)$. This edge has 2 interior points. Hence the divisor $D_5$ corresponding to $\nu_5 = (1,1,0,0)$ consists of three rational curves $D_{5}^\alpha$ which each meet $D_1$ in a single point. A basis of $Pic(S)$ is given by $D_{5}^1,D_{5}^2,D_{5}^3,D_1$ which have inner form
\begin{equation}
Pic(S) \sim  \left(\begin{array}{rrrr} 
-2 & 0 & 0 & 1 \\
0 & -2 & 0 & 1 \\
0 & 0 & -2 & 1 \\
1 & 1 & 1 & 0
\end{array}\right) \, .
\end{equation}
Note that only $D_5 = D_5^1 + D_5^2 + D_5^3$ and $D_1$ are realized as toric divisors, so that $Pic_{tor}(S)$ is two-dimensional and has inner form
\begin{equation}
Pic_{tor}(S) \sim  \left(\begin{array}{rr}
-6 & 3 \\
3 & 0 
\end{array}\right)\, .
\end{equation}
As we have considered the trivial top over $\Delta^\circ_F$ above, for which all faces are non-vertical, only 
$Pic_{tor}(S_0)$ is in the image of $\rho$, so that $rk(N)=2$. Furthermore, $rk(K)=0$ as all divisor classes except $[S_0]$ restrict non-trivially 
to $S_0$. Geometrically, the $K3$ fibration $Z$ has a non-trivial monodromy acting on $Pic(S)$ which only leaves $D_5$ and $D_1$ invariant. Correspondingly, only those two become become divisors on $Z$.

We can change the top $\Diamond^\circ$ to $\Diamond^{'\circ}$ by using as a vertex $(1,1,0,1)$ above $F$ instead of $(0,0,0,1)$. This turns the face which contains $(1,1,0,0)$ in its boundary into a vertical face and we hence expect all of the divisors $D_5^\alpha$ to give rise to divisors of $Z'$. Correspondingly, \eqref{eq:hodgenumbersZ} now tells  us that 
\begin{equation}
 h^{1,1}(Z') = 5 \hspace{1cm} h^{2,1}(Z) = 42\, ,
\end{equation}
and we find that $N = Pic(S)$, so that $rk(N) = 4$ while keeping $K = 0$. 

\subsection{A Very Large Top}\label{sect:extrex}

Finally, let us consider a rather extreme example. There is a unique polytope $\Delta^\circ$ in the Kreuzer-Skarke list which has the maximal possible number of lattice points of any reflexive polytope, $680$. The Calabi-Yau manifold $X_{(\Delta, \Delta^\circ)}$ has the largest Euler characteristic of any toric hypersurface Calabi-Yau threefold, $960$, and is the mirror to a generic elliptic fibration over the Hirzebruch surface $\mathbb{F}_{12}$.  

As discussed in \cite{Candelas:2012uu}, we can cut $\Delta^\circ$ in half to find two (isomorphic) projecting tops $\Diamond^\circ$. They are the largest projecting tops over $\Delta^\circ_F$ corresponding to an elliptic $K3$ surface with one $II^*$ fibre. The vertices of $\Diamond^\circ$ and its dual are
\begin{equation}
\Diamond^\circ \sim \left(\begin{array}{rrrrr}
-1 & 0 & 0 & 6 & 6 \\
2 & -1 & 0 & 2 & 2 \\
3 & 0 & -1 & 3 & 3 \\
0 & 0 & 0 & 0 & 42
\end{array}\right)\, , \hspace{1cm}
\Diamond\sim\left(\begin{array}{rrrrr}
-1 & 0 & 6 & 6 & 0 \\
1 & -2 & 1 & 1 & 1 \\
1 & 1 & 1 & 1 & -1 \\
0 & 0 & 0 & -1 & 0
\end{array}\right)\, .
\end{equation}
The generic fibre $S$ has $Pic(S) = Pic_{tor}(S)= U \oplus (-E_8)$, so that $N = U \oplus (-E_8)$. Evaluating \eqref{eq:hodgenumbersZ} reveals that
\begin{equation}
 h^{1,1}(Z) = 251\,, \hspace{1cm} h^{2,1}(Z) = 0 \, ,
\end{equation}
so that $rk(K) = 240$. 

We may construct an extraordinary $G_2$ manifold $X_{Z,Z}$ by gluing two building blocks $Z_\pm$ both isomorphic to the $Z$ just constructed (times $S^1$) such that $N_+ \cap N_- = 0$ \footnote{Strictly speaking, this requires proving that we can choose the complex structure of the algebraic family such that the matching \eqref{eq:hkrot} can be satisfied, i.e. $Im(\Omega_\pm) \subset U$ and $Re(\Omega_\pm) \subset U \oplus (-E_8)$. In this case we can choose a complex structure and a K\"ahler form such that there exists a (rather trivial) lattice isometry inducing \eqref{eq:hkrot}. For this example, we unfortunately cannot use the surjectivity of \eqref{eq:modhypvslpk3}, as the polytope $\Delta^\circ_F$ has interior points in 2-dimensional faces, so that the ambient space of the fibre is not semi Fano.}.  This means that $\Lambda/(N_+ + N_-) = U$ and $N_\pm \cap T_\mp = U \oplus (-E_8)$. By \eqref{eq:bettiG2}, such a $G_2$ manifold will have Betti numbers 
\begin{equation}
b_2(X_{Z,Z}) = b_5(X_{Z,Z}) = 480 \, , \hspace{1cm} b_3(X_{Z,Z}) = b_4(X_{Z,Z}) = 503 \, .
\end{equation}
An M-Theory compactification on $X_{Z,Z}$ will hence have gauge group $U(1)^{480}$ ! To our knowledge, this would make $X_{Z,Z}$ the $G_2$ manifold with the largest Betti numbers constructed so far. 

Reversing the roles of $\Diamond^\circ$ and $\Diamond$ yields a building block $\hat{Z}$ with the same $K3$ fibre ($Pic(S) = N = U \oplus (-E_8)$) but with
\begin{equation}
 h^{1,1}(\hat{Z}) = 11\,, \hspace{1cm} h^{2,1}(\hat{Z}) = 240 \, ,
\end{equation}
so that now $rk(K)=0$. 

Orthogonally glueing two copies of $\hat{Z}$ (times $S^1$) to form a $G_2$ manifold $X_{\hat{Z},\hat{Z}}$ or orthogonally 
gluing one copy of $Z$ with one copy of $\hat{Z}$ to form a $G_2$ manifold $X_{Z,\hat{Z}}$ then results in the Betti numbers
\begin{equation}
\begin{aligned}
b_2 (X_{Z,\hat{Z}}) = 240 \, , \hspace{1cm} b_3(X_{Z,\hat{Z}}) = 743 \\
b_2(X_{\hat{Z},\hat{Z}}) = 0 \, , \hspace{1cm} b_3(X_{\hat{Z},\hat{Z}}) = 983 \, .
\end{aligned}
\end{equation}
We cannot help but wonder if there are singular transitions in M-Theory connecting the compactifications on $X_{Z,Z}$, $X_{\hat{Z},\hat{Z}}$ and $X_{Z,\hat{Z}}$. Investigating the associated physics promises to be very interesting.

\section{Discussion and Outlook}

In this work we have demonstrated that four-dimensional projecting tops can be straight-forwardly used to manufacture the building blocks used to construct $G_2$ manifolds as twisted connected sums (TCS). This is not at all unexpected, as such tops capture the geometry of `half' a $K3$-fibred Calabi-Yau threefold. Conveniently, a great deal of information on the geometry of both a generic $K3$ fibre and its degeneration is captured by the combinatorics of the top. In particular, it is possible to combinatorially determine the relevant topological data: the Hodge numbers of the building block, the group $N$ and the rank of $K$. While these results are primarily of a mathematical nature, it is hoped that they pave the way for a better understanding of physics associated with compactifications of M-Theory on $G_2$ manifolds.

Realizing $K3$ fibres embedded in toric varieties in order to construct building blocks is nothing new and has been already explored in \cite{corti_weakfano}. The description in this paper, however, allows for an elegant characterization of $K3$ fibrations with degenerate fibres. The usefulness of this is twofold. First of all, it allows a constructive control over the Hodge numbers of building blocks and hence the Betti numbers of resulting $G_2$ manifolds. Furthermore, all of this is done in a language familiar to string theorists. This should further strengthen the bridge from complex algebraic geometry to $G_2$ manifolds provided by the TCS construction, in particular in applications to physics. Secondly, as is well-known from the description of Calabi-Yau manifolds, polytopes anticipate possible singular transitions. Presently, this is only clear on the level of the building blocks and more work is needed in order to formulate clear criteria when such a transition is also possible in $G_2$ moduli space. 

The construction of $K3$ fibred threefolds with $c_1 = [fibre]$ from tops we have presented is easily generalized to arbitrary dimensions. Here, the result is a Calabi-Yau n-fold fibred algebraic $n+1$-fold with $c_1 = [fibre]$, obtained from a $n+2$-dimensional top. As already remarked, this construction yields rational elliptic surfaces for $n=1$. As is well-known, there is a semi-stable degeneration of a family of $K3$ surfaces into a pair of rational elliptic surfaces. When the $K3$ surface is described as a toric hypersurface and the corresponding polytope can be disassembled into a top and bottom, our construction yields an explicit realization of the two rational elliptic surfaces. It seems likely that similar semi-stable degenerations of Calabi-Yau n-folds are encoded in projecting tops for arbitrary dimensions. From this point of view, building blocks can be thought of as three-dimensional generalizations of rational elliptic surfaces. 

From the point of view of physics, the most interesting aspect of $G_2$ manifolds lies in their possible degenerations. As already remarked, the description of building blocks given here allows for a straightforward construction of singular building blocks. It remains to be seen under which circumstances such singularities can also be reached in the moduli space of the associated $G_2$ manifolds. Clearly, this necessitates a detailed study of (co)-associative submanifolds, which form the calibrated cycles of a $G_2$ manifold. There exist results on how such submanifolds arise from the point of view of the TCS construction \cite{corti_g2}, and many interesting ideas for how this can be employed to degenerate the associated $G_2$ manifolds have been discussed in \cite{Halverson:2014tya,Halverson:2015vta}. 

While questions about existence of the gluing morphism can be cast in the language of lattice embeddings using the Torelli theorem\footnote{This statement only holds modulo the fact that the moduli space of a toric hypersurface is not necessarily the same as that of the associated lattice polarized family.}, it remains an open problem to classify all possibilities. As we have seen, both the Picard lattice and the lattice $N$ can be 
readily determined from the polytope data. It remains to be seen whether this can be used to describe at least a well-defined class of all possible matchings for two given pairs.  

Finally, it should be noted that the TCS construction describes a $G_2$ manifold as a (non-holomorphic) $K3$ fibration over $S^3$ base, which ties in nicely with the M-Theory/heterotic duality and the SYZ fibration. If will be interesting to see if this can be exploited to learn about either side of the duality map.

\section*{Acknowledgements}

I wish to thank Philip Candelas, Xenia de la Ossa, Jim Halverson, Andre Lukas, Sakura Sch\"afer-Nameki and Taizan Watari for helpful discussions, inspiration and support. This work was supported by STFC grant ST/L000474/1 and EPSCR grant EP/J010790/1. This work was in part performed at the Aspen Center for Physics, which is supported by National Science Foundation grant PHY-1066293

\appendix

\section{Lattices, Mirror Symmetry for K3 Surfaces and Reflexive Polytopes}\label{sect:background}

\subsection{Lattices}

In this section a few helpful facts on lattices are reviewed. A good reference for this is given by \cite{Nikulin}.

A lattice $L$ is a free abelian group of finite rank, so that $L\cong \Z^n$. It is endowed with an inner form $\langle \cdot, \cdot \rangle$ and 
the numbers $r_{\pm}$ of positive/negative eigenvalues are called its signature. Often it it convenient to think of a lattice as being obtained by forming arbitrary linear combinations, with integer coefficients $n_i$, of a finite set of generators $v_i$ in $\R^{r_+,r_-}$ so that for any $l\in L$
\begin{equation}
 l = \sum_i n_i v_i \, .
\end{equation}
In this way, a lattice inherits its inner form from $\R^{r_+,r_-}$. The dual lattice $L^*$ contains all vectors which have an integer inner product with all elements of $L$. 

A lattice is called even if all squares $\langle l,l\rangle$ of its elements are even integers.
An unimodular lattice $L$ is a lattice such that $L^* \cong L$. Even unimodular lattices only exist when $r_+ - r_- = 0 \,\mbox{mod}\, 8$, and they are unique (up to isomorphism) if both $r_-$ and $r_+$ are non-vanishing. In this case 
\begin{equation}
 L \cong \pm E_8^{a} \oplus U^{b} \, ,
\end{equation}
for some integers $a$ and $b$. Here, $\pm E_8$ is ($\pm$) the root lattice of $E_8$ and $U$ is the hyperbolic lattice with inner form
\begin{equation}
 \left(\begin{array}{cc}
  0 & 1 \\
  1 & 0
 \end{array}\right)\, .
\end{equation}

For non-unimodular lattices, there is a nontrivial quotient $G_L = L^*/L$ which is a finite abelian group. On $G_L$, there is an inner quadratic form
$q_L$ and a bilinear form $b_L$ defined by
\begin{equation}
 \begin{aligned}
  q_L: & x \rightarrow \langle x, x\rangle \\
  b_L:& (x,y)\rightarrow \frac12 \left(q_L(x+y)-q_L(x) -q_L(y) \right) \, ,
 \end{aligned}
\end{equation}
$q_L$ and $b_L$ take values in $\Q/2\Z$ and $\Q/\Z$, respectively.

An embedding $N \hookrightarrow L$ of lattices is called primitive if the quotient is torsion-free. This is equivalent to saying that
there is no lattice element $n$ in $N$ such that $n/m  \in L$ but not in $N$ for any integer $m > 1$. Note that this means that primitive embeddings are transitive, if $N \hookrightarrow M$ is primitive and $M \hookrightarrow L$ is primitive, then so is $N \hookrightarrow L$.

For a primitive embedding of a lattice $N$ into 
an even unimodular lattice $L$, $N$ and its orthogonal complement $N^\perp$ in $L$ share the same discriminant form (up to a sign)
\begin{equation}\label{eq:disrcperp}
\begin{aligned}
  G_N &\cong G_{N^\perp} \\ 
 q_N &\cong - q_{N^\perp} \, .
\end{aligned}
 \end{equation}
This statement has a converse, i.e. for two lattices $N$ and $M$ there exists a primitive embedding into an even unimodular lattice $L$ such that $N= M^\perp$ in $L$ (and vice versa) if their discriminant forms satisfy \eqref{eq:disrcperp} and if their ranks and signatures add up in the obvious way.

\subsection{Lattice Polarized K3 Surfaces and Mirror Symmetry}
\label{sect:mirk3}

Mirror symmetry for $K3$ surfaces first appeared in \cite{Pinkhamk3,DGNI,DG,Nikulin} as an observation in mathematics related to Arnold's strange duality and was first discussed in \cite{Aspinwall:1994rg} in its physical incarnation, see also \cite{1995alg.geom..2005D}.

A $K3$ surface is the unique simply-connected Calabi-Yau manifold in two complex dimensions. Its only non-trivial Hodge number (besides $h^{0,0}=h^{2,2}=h^{2,0}=h^{0,2}=1$) is $h^{1,1}=20$.
Due to Poincar\'e duality and the relation
\begin{equation}
\chi(C)  = 2g-2 = - C \cdot C\, ,
\end{equation}
for every (irreducible) curve\footnote{Of course, not every element in $H^2(K3,\Z)$ corresponds to an irreducible curve, but $H^2(K3,\Z)$ is generated by such curves.}, $H^2(K3,\Z)$ becomes an even self-dual lattice. Its signature is $(3,19)$, so that $H^2(S,\mathbb{Z}) \cong (-E_8)^{\oplus 2}\oplus U^{\oplus 3} \equiv \Lambda$.

An isometry $h:H^2(S,\mathbb{Z})\mapsto \Lambda$ is called a marking. Using a 
marking, the period map identifies a point in $\Lambda \otimes \C$ as the image of 
$\Omega$.

A family of lattice polarized K3 surfaces with polarizing lattice $L$ is a 
family for which the holomorphic two-form $\Omega$ is perpendicular (under the natural inner form on $H^2$) to $L$. 
The period domain of a lattice polarized K3 surface is given by
\begin{equation}\label{eq:perioddomain}
P_L = \{\Omega\,\, |\,\, \Omega \cdot \Omega = 0\, , \Omega \wedge \bar{\Omega} 
> 0\, ,\Omega\cdot L = 0\} 
\end{equation}
The period map is surjective onto the period domain and the moduli space of a lattice polarized K3 surface is given by 
\begin{equation}
P_L/\Gamma 
\end{equation}
where $\Gamma$ is the group of automorphisms of $\Lambda$.

One incarnation of mirror symmetry for K3 surfaces associates families of lattice polarized K3 surfaces. 
This topic goes back a long way in mathematics \cite{1995alg.geom..2005D} and exists in several versions. In physics \cite{Aspinwall:1994rg} the objects we should consider as mirror pairs are pairs of lattice polarized K3 surfaces with 
lattices $N$ and $N'$ such that there is a primitive embedding
\begin{equation}\label{eq:k3mirror}
 N \oplus N' \oplus U \hookrightarrow \Lambda
\end{equation}

From the embedding \eqref{eq:k3mirror} we learn that the discriminant forms of $N$ and $N'$ must agree as in \eqref{eq:disrcperp}. Conversely, for any lattice $N$ which can be primitively embedded into $N \hookrightarrow U^{\oplus 2}\oplus (-E_8)^{\oplus 2}$ there is an associated mirror pair of lattice polarized families of K3 surfaces.

\subsection{K3 surfaces from Reflexive Polyhedra}

Such a correspondence is indeed realized \cite{Rohsiepe:2004st} by using Batyrev's construction 
of dual reflexive polytopes \cite{1993alg.geom.10003B}. Starting 
from a pair of dual reflexive polytopes $\Delta^\circ$ and $\Delta$,
we can define a toric variety $\P_\Sigma$ by constructing a fan $\Sigma$ from a (regular, fine, star)
triangulation\footnote{Although 3-dimensional polytopes do not have a unique triangulation, the strata which have a non-trivial restriction to a Calabi-Yau hypersurface are unique.} of $\Delta^\circ$. A generic Calabi-Yau (K3) hypersurface $X_{(\Delta,\Delta^\circ)}$ is then defined by the equation 
\be\label{eq:k3monos}
\sum_m c_m \prod_{\nu_i \in \Sigma(1)} z_i^{\langle m, \nu_i \rangle+1} = 0
\ee
with generic coefficients $c_m \in \C$ for every integral point $m$ on $\Delta$. 

\subsubsection{\texorpdfstring{$\mbox{Pic}_{tor}(X_{(\Delta,\Delta^\circ)})$}{Lg}}

The toric divisors $D_i$ give rise to a sublattice $Pic_{tor}(X_{(\Delta,\Delta^\circ)})$ of the Picard 
lattice of $X_{(\Delta,\Delta^\circ)}$. Let us describe this sublattice in some
more detail. For every ray in $\Sigma(1)$ which is not interior to a 
two-dimensional face of $\Delta^\circ$, there is a corresponding 
generator in $Pic_{tor}(X_{(\Delta,\Delta^\circ)})$. Due to the linear relations
\begin{equation}\label{eq:linrel}
 \langle \nu_i, m \rangle D_i = 0 \,\,\forall m \in {\bf M}
\end{equation}
$Pic_{tor}(X_{(\Delta,\Delta^\circ)})$ becomes a sublattice (of codimension 3) of 
$\mathbb{Z}^{|\Sigma(1) \notin \Theta^{\circ[2]}|}$. 

The inner form on $Pic_{tor}(X_{(\Delta,\Delta^\circ)})$
is determined by combinatorial data and can be computed using the theory of 
\cite{DK}. Every $\nu_i$ corresponding 
to a vertex of $\Delta^\circ$ is a curve of genus $\ell^*(\Theta^{[2]})$, where 
$\Theta^{[2]}$ is the dual 2-dimensional 
face on $\Delta$ and $\ell^*$ counts interior lattice points. 
Hence $D_i \cdot D_i = 2\ell^*(\Theta^{[2]})-2$ for any 
vertex $\nu_i$. A toric divisor $D_i$ corresponding 
to a point interior to a 1-dimensional face $\Theta^{\circ [1]}$ decomposes 
into $1+\ell^*(\Theta^{[1]})$ $\P^1$s when
restricted to the K3 hypersurface. Only their sum is contained in $\mbox{Pic}_{tor}(X_{(\Delta,\Delta^\circ)})$.
These $\P^1$s are not mutually intersecting, 
so that $D_i \cdot D_i = -2 (1+\ell^*(\Theta^{[1]}))$
for $\nu_i$ an interior point of a 1-dimensional face $\Theta^{\circ [1]}$. 
Finally, the intersecting between two toric divisors 
$D_i$ and $D_j$ is non-zero only if they are neighbours along a common 
1-dimensional face $\Theta^{\circ [1]}$. In this case
$D_i \cdot D_j = 1+\ell^*(\Theta^{[1]})$, where $\Theta^{[1]}$ is the dual 
1-dimensional face on $\Delta$. This finishes the description of
$Pic_{tor}(X_{(\Delta,\Delta^\circ)})$ which has rank
\begin{equation}
|Pic_{tor}(X_{(\Delta,\Delta^\circ)})| = \ell(\Delta^\circ) - 1 - 3 -\sum_{\Theta^{\circ 
2}}\ell^*(\Theta^{\circ 2}) = \ell^1(\Delta_F^\circ) - 3 \, ,
\end{equation}
where $\ell^1$ counts lattice points on the one-skeleton. 

\subsubsection{\texorpdfstring{$\mbox{Pic}(X_{(\Delta,\Delta^\circ)})$}{Lg}}

Let us now describe the polarizing lattice of the family of K3 hypersurfaces $X$ in $\P_\Sigma$. As we have already mentioned, toric divisors corresponding to points interior to 1-dimensional faces of $\Delta^\circ$ can become reducible when
restricted to $X$, so that $\mbox{Pic}(X_{(\Delta,\Delta^\circ)})$ is, in general, larger than
$\mbox{Pic}_{tor}(X_{(\Delta,\Delta^\circ)})$. Again, using the stratification associated with the ambient toric variety $\P_\Sigma$
and the theory of \cite{DK} gives a detailed picture of divisors on $X_{(\Delta,\Delta^\circ)}$ appearing as components of the restriction of 
toric divisors. Note that this method does not guarantee that we have succesfully determined the whole Picard lattice of a generic member of the algebraic family of $K3$ surfaces we are considering (although the consideration of mirror pairs of $K3$ surfaces suggests this is the case), see \cite{grassi_bruzzo} for a more rigourous treatment. Keeping in mind this cautionairy remark, we continue to denote the lattice obtained by the stratification method by $\mbox{Pic}(X_{(\Delta,\Delta^\circ)})$ throughout this article. Note that we only require the image of the Picard group of the generic fibre under the restriction map for the present work, see Sections \ref{sect:defsbuildbocks} and \ref{sect:NandK}, which we can confidently identify.

As divisors corresponding to vertices are irreducible curves, we still associate a divisor $D_j$ to every 
vertex $\nu_j$. For every point $\nu_i$ interior to a 1-dimensional face $\Theta^{\circ [1]}$ there are $1+\ell^*(\Theta^{[1]})$ 
divisors $D_i^\alpha$, $\alpha = 1,\cdots,1+\ell^*(\Theta^{[1]})$.
They satisfy $D_i = \sum D_i^\alpha$. Again, the linear relations are given by \eqref{eq:linrel}.

The inner form is given by 
\be
\begin{aligned}
D_i \cdot D_i &=  2\ell^*(\Theta^{[2]})-2 \,\,\mbox{for}\,\, \nu_i\,\, \mbox{a 
vertex} \\
D_i^\alpha \cdot D_i^\beta &= -2 \delta^{\alpha \beta} \\
D_j^\alpha \cdot D_i^\beta &= \delta^{\alpha \beta}  \,\, \mbox{if}\,\, \nu_i, \nu_j \,\, 
\mbox{are neighbours along a 1D face } \\
& = 0\,\, \mbox{otherwise} \\
D_j^\alpha \cdot D_i &=  1 \,\, \mbox{if}\,\, \nu_i, \nu_j \,\, \mbox{are 
neighbours along a 1D face and}\,\, \nu_i\,\,\mbox{is a vertex} \\
D_i \cdot D_j& = 1 + \ell^*(\Theta^{[1]}) \,\, \mbox{if}\,\, \nu_i,\nu_j \,\, 
\mbox{vertices connected along face} \,\, \Theta^{\circ [1]}
\end{aligned}
\ee

Note that we may think of $\mbox{Pic}_{tor}(X_{(\Delta,\Delta^\circ)})$ as a sublattice of $\mbox{Pic}(X_{(\Delta,\Delta^\circ)})$ such that
\begin{equation}\label{eq:picvspictorL}
\mbox{Pic}(X_{(\Delta,\Delta^\circ)})/ \mbox{Pic}_{tor}(X_{(\Delta,\Delta^\circ)}) = \bigoplus_i L_{i}
\end{equation}
with $L_i$ being the lattice associated to a dual pair of one-dimensional faces $(\Theta^{[1]}_i,\Theta^{\circ[1]}_i)$:
\begin{equation}
L_i =  A_{\ell^*(\Theta^{\circ[1]}_i)}^{\oplus \ell^*(\Theta^{[1]}_i)}
\end{equation}

\subsection{Mirror Symmetry and Reflexive Three-Dimensional Polyhedra}

From the discussion above it follows that 
\begin{equation}\label{eq:pick3}
|Pic(X_{(\Delta,\Delta^\circ)})| = \ell(\Delta^\circ) - 1 - 3 -\sum_{\Theta^{\circ 
[2]}}\ell^*(\Theta^{\circ 2}) + \sum_{\Theta^{\circ [1]}} \ell^*(\Theta^{\circ 
[1]})\ell^*(\Theta^{[1]})
\end{equation}
gives us the rank of the polarizing lattice (i.e. Picard group of a generic 
member) of the family $X_{(\Delta^\circ,\Delta)}$ described above.
Applying this formula to a pair of dual reflexive polytopes naively leads to 
contradiction with \eqref{eq:k3mirror} as 
\begin{equation}
|Pic(X_{(\Delta,\Delta^\circ)})| + |Pic(X_{(\Delta^\circ,\Delta)})| = 20 +  \sum_{\Theta^{\circ [1]}} \ell^*(\Theta^{\circ 
[1]})\ell^*(\Theta^{[1]}) \, ,
\end{equation}
i.e. the dimension is too large by the value of the correction term in 
\eqref{eq:pick3}.

The dimensions do work out if we consider mirror families with lattice polarizations
\be\label{eq:nnp}
\begin{aligned}
N &= Pic(X_{(\Delta,\Delta^\circ)}) \\
N'&= Pic_{tor}(X_{(\Delta^\circ,\Delta)})
\end{aligned}
\ee
In fact, as has been explicitly verified in \cite{Rohsiepe:2004st} (and repeated by the present author), the 
discriminant forms on $Pic(X_{(\Delta,\Delta^\circ)})$ and $Pic_{tor}(X_{(\Delta^\circ,\Delta)})$ agree for a pair of 
reflexive polytopes, so that a primitive embedding \eqref{eq:k3mirror} exists and we have found a mirror pair.

The reason for this seemingly asymmetric choice is that the correction term $\sum_{\Theta^{\circ [1]}} 
\ell^*(\Theta^{\circ [1]})\ell^*(\Theta^{[1]})$ 
counts non-toric divisors \emph{and} non-polynomial deformations of $X_{(\Delta,\Delta^\circ)}$ at the same time. The two are mutually 
exclusive, however: if one constructs $X_{(\Delta,\Delta^\circ)}$ as a hypersurface by summing the monomials \eqref{eq:k3monos} all divisors which we have 
described above as $Pic(X_{(\Delta,\Delta^\circ)})$ are sitting in the Picard lattice, including the non-toric ones, \eqref{eq:picvspictorL}.
Furthermore there are $|Pic_{tor}(X_{(\Delta^\circ,\Delta)})|$ polynomial deformations which 
preserve these divisors\footnote{This can be seen by counting the number of monomials and subtracting the rank of the group
of automorphisms of $\P_\Sigma$.}. As soon as we consider non-polynomial deformations, cycles which are in in $Pic(X_{(\Delta,\Delta^\circ)})$ but not in 
$Pic_{tor}(X_{(\Delta,\Delta^\circ)})$ cease to be purely of Hodge type $(1,1)$ so that they are no longer in the Picard lattice and the 
corresponding family is lattice polarized with polarizing lattice $Pic_{tor}(X_{(\Delta,\Delta^\circ)})$ instead of $Pic(X_{(\Delta,\Delta^\circ)})$. 

Given a pair of reflexive polytopes, we hence have to make a choice as to how 
to `distribute' the correction term $\sum_{\Theta^{\circ [1]}} \ell^*(\Theta^{\circ [1]})\ell^*(\Theta^{[1]})$ between 
$N$ and $N'$. The easiest choice is the one described in \eqref{eq:nnp}, which is what we adopt in the present work. In general, there are other choices 
if $\sum_{\Theta^{\circ [1]}} \ell^*(\Theta^{\circ [1]})\ell^*(\Theta^{[1]}) > 0$. Due to this subtlety, a pair of reflexive 3-dimensional polytopes generally gives rise to \emph{several} mirror pairs of families of lattice polarized K3 surfaces.

\section{Toric Calabi-Yau Hypersurfaces and Toric Stratification}
\label{app:hodge_strat}

As an application in a more familiar setting, we use the technique of stratification employed for building blocks in Section \ref{sect:NandK} to compute the Hodge numbers of toric Calabi-Yau manifolds in this appendix. In this case, the ${\bf M}$ lattice polytope $\Delta$ is reflexive, i.e. there is a lattice polytope $\Delta^\circ$ such that
\begin{equation}
 \langle \Delta, \Delta^\circ \rangle \geq -1 \, ,
\end{equation}
and the normal fan $\Sigma_n(\Delta)$ is identical to the fan over the faces of $\Delta^\circ$, $\Sigma_f(\Delta^\circ)$ \cite{1993alg.geom.10003B}. In this correspondence we may identify
\begin{equation}\label{eq:dualfacesandcones}
\sigma_f(\Theta^{\circ [k]}) = \sigma_n(\Theta^{[n-k-1]}) \, , 
\end{equation}
for a dual pair of faces satisfying
\begin{equation}
\langle \Theta^{[n-k-1]},\Theta^{\circ [k]} \rangle = -1  \, .
\end{equation}

\subsection{Hodge Numbers of Toric Calabi-Yau Hypersurface}

To compute the Hodge numbers of Calabi-Yau threefold hypersurfaces, let us first note that they enjoy a stratification
\begin{align}
 Z = Z_{\Delta} \amalg Z_{\Theta^{[3]}} \amalg Z_{\Theta^{[2]}} \times [ \underbrace{\sum \C^*}_{\mbox{points on } \Theta^{\circ [1]}} 
 + \underbrace{\sum pt}_{\mbox{1-simplices on } \Theta^{\circ [1]}} ]\nn \\ \amalg Z_{\Theta^{[1]}} \times [ \underbrace{\sum (\C^*)^2}_{\mbox{points on } \Theta^{\circ [2]}}  + \underbrace{\sum \C^*}_{\mbox{1-simplices on } \Theta^{\circ [2]}} + \underbrace{\sum pt}_{\mbox{2-simplices on } \Theta^{\circ [2]}} ]
\end{align}
after we have resolved the variety $\P_{\Sigma_n(\Delta)}$ using all lattice points on $\Delta^\circ$. Here we have used that due to \eqref{eq:dualfacesandcones}, the location of cones of $\Sigma$ within the normal fan $\Sigma_n(\Delta)$ is identical to the location of the corresponding simplex within faces of $\Delta^\circ$.

We may now simply compute $h^{1,1}(Z)=h^{2,2}(Z) = e^{2,2}(Z)$ by summing the numbers $e^{2,2}$ of the individual strata. In the expressions below, $\Theta^{\circ[k]}$ of $\Delta^\circ$ is considered to be the dual to the face $\Theta^{[4-k-1]}$ being summed over.
\begin{align}\label{eq:h11X}
 h^{1,1}(Z) & = e^{2,2}(Z_{\Delta}) + \sum_{\Theta^{[3]}<\Delta} e^{2,2}(\Theta^{[3]}) + \sum_{\Theta^{[2]}<\Delta}e^{1,1}(\Theta^{[2]})\ell^*(\Theta^{\circ[1]}) \nn \\
& \hspace{1cm} + \sum_{\Theta^{[1]}<\Delta}e^{0,0}(\Theta^{[1]})\,\cdot\, \ell^*(\Theta^{\circ[2]})\, \cdot\,e^{2,2}((\C^*)^2) \nn \\
 & = -4 + \sum_{\Theta^{[3]}<\Delta} 1 + \sum_{\Theta^{[2]}<\Delta} \ell^*(\Theta^{\circ[1]}) + \sum_{\Theta^{[1]}<\Delta}(\ell^*(\Theta^{[1]})+1)\ell^*(\Theta^{\circ[2]}) \nn \\
 & = \ell(\Delta) - 5 - \sum_{\Theta^{\circ [3]}<\Delta^\circ}\ell^*(\Theta^{\circ [3]})  + \sum_{\Theta^{\circ [2]}<\Delta^\circ}\ell^*(\Theta^{[1]})\ell^*(\Theta^{\circ[2]}) \, .
\end{align}
In the last line we have used that we may associate a unique vertex to each face $\Theta^{[3]}$ and that we can decompose all points on $\Delta^\circ$
according to the face containing them in their relative interior.

Similarly, we can compute $h^{2,1}(Z) = -e^{2,1}(Z)$. As $e^{2,1}(Z_{\Theta^{[3]}}) = 0$, we find 
\begin{align}\label{eq:h21X}
 e^{2,1}(Z) & =  e^{2,1}(Z_{\Delta}) + \sum_{\Theta^{[2]} < \Delta} e^{1,0}(Z_\Theta^{[2]})\,\cdot\, \ell^*(\Theta^{\circ [1]})\,\cdot\,e^{1,1}(\C^*) \\
 & = \ell^3(\Delta) - \ell^2(\Delta) - \varphi_2(\Delta) -  \sum_{\Theta^{[2]}<\Delta} \ell^*(\Theta^{[2]})\ell^*(\Theta^{\circ [1]}) \\
 & = \sum_{\Theta^{[3]}<\Delta} \ell^*(\Theta^{[3]})  +5 - \ell(\Delta) - \sum_{\Theta^{[2]}<\Delta} \ell^*(\Theta^{[2]})\ell^*(\Theta^{\circ [1]})  \, .
\end{align}
Here, we have used that because $\ell^*(2\Delta) =  \ell(\Delta) $ and $ \ell^*(\Delta)=1$, the function $\varphi_2(\Delta)$ is simply given by 
\begin{align}
 \varphi_2(\Delta) & =  \sum_{j=1}^2(-1)^j\binom{5}{2-j}\ell^*(j \Delta) \\
 & = - 5 \ell^*(\Delta) + \ell^*(2 \Delta) \\
 & = -5 + \ell(\Delta) \, .
\end{align}
Hence we end up with the final result
\begin{equation}
 h^{2,1}(Z) = \ell(\Delta) - 5 - \sum_{\Theta^{[3]}<\Delta} \ell^*(\Theta^{[3]}) + \sum_{\Theta^{[2]}<\Delta} \ell^*(\Theta^{[2]})\ell^*(\Theta^{\circ [1]}) \, . 
\end{equation}

\subsection{Topology of Divisors}

In this section we derive combinatorial formulas for the Hodge numbers $h^{0,i}$ for toric divisors of Calabi-Yau threefolds.
The Hodge numbers $h^{1,1}$ of toric divisors depend on the triangulation data and are captured by slightly more complicated expression, which can however also easily be obtained from the methods used. 

\subsubsection{Vertices}

A divisor $Y_i$ for which $\nu_i$ is a vertex, dual to $\Theta^{[3]}$, has a stratification
\begin{equation}\label{eq:stratYvertex}
Y_i = Z_{\Theta^{[3]}}  \amalg_{\Theta^{[2]} < \Theta^{[3]}} Z_{\Theta^{[2]}}\times \left( pt \right) 
\amalg_{\Theta^{[1]} < \Theta^{[3]}}  Z_{\Theta^{[1]}}\times \left(\sum \C^* \amalg \sum pt \right) \, .
\end{equation}
Here the $pt$ multiplying $ Z_{\Theta^{[2]}}$ originates from the unique 1-simplex on $\Theta^{\circ[1]}$ dual to $\Theta^{[2]}$
($k=2,l=2$ in \eqref{eq:masterstratres}). The $\C^*$s multiplying $Z_{\Theta^{[1]}}$ originate from 1-simplices and the $pt$s from 2-simplices 
on each face $\Theta^{\circ[2]}$ dual to $\Theta^{[1]}$.

With the stratification \eqref{eq:stratYvertex} at hand, we can start computing the Hodge numbers. As $Z_{\Theta^{[3]}}$ is an irreducible 
open complex surface, also the divisor $Y_i$ is irreducible. For $h^{1,0}(Y_i)$, only the first two strata contribute and we find
\begin{equation}
\begin{aligned}
h^{1,0}(Y_i) & = -e^{1,0}(Y_i) \\
 & = -\left(e^{1,0}(Z_{\Theta^{[3]}}) + \sum_{\Theta^{[2]} < \Theta^{[3]}}e^{1,0}(Z_{\Theta^{[2]}}) \right) \\
 & = -\left(\sum_{\Theta^{[2]} < \Theta^{[3]}}\ell^*(\Theta^{[2]}) - \sum_{\Theta^{[2]}< \Theta^{[3]}}\ell^*(\Theta^{[2]}) \right) 
 & = 0
\end{aligned}
\end{equation}
For $h^{2,0}$, only the first stratum in  \eqref{eq:stratYvertex} contributes and we find
\begin{equation}
 h^{2,0}(Y_i) = e^{2,0}(Z_{\Theta^{[3]}})  = \ell^*(Z_{\Theta^{[3]}}) \, .
\end{equation}

\subsubsection{Divisors Interior to 1D Faces of \texorpdfstring{$\Delta^\circ$}{Lg}}

Divisors originating from points $\nu_i$ interior to edges $\Theta^{\circ[1]}$ of $\Delta^\circ$ dual to
two-dimensional faces $\Theta^{[2]}$ of $\Delta$ have a stratification
\begin{equation}\label{eq:stratYedge}
Y_i = Z_{\Theta^{[2]}} \times \left(\C^* + 2pts \right) 
\amalg_{\Theta^{[1]}<\Theta^{[2]}} Z_{\Theta^{[1]}}\left(\sum \C^* + \sum pt \right) \, .
\end{equation}
Here, the $\C^*$ multiplying $Z_{\Theta^{[2]}}$ is due to the point itself ($k=2,l=1$), whereas the 2 points correspond
to the two 1-simplices on $\Theta^{\circ[1]}$ containing $\nu_i$ ($k=2,l=2$). Each $\C^*$ multiplying $Z_{\Theta^{[1]}}$
corresponds to a 1-simplex containing $\nu_i$ which is interior to the dual $\Theta^{\circ [2]}$ and each $pt$ corresponds
to a 2-simplex containing $\nu_i$ which is interior to the dual $\Theta^{\circ [2]}$. 

Again, $h^{0,0}(Y_i)=1$ as $Y_i$ is irreducible. The computation for $h^{1,0}$ now becomes
\begin{equation}
 \begin{aligned}
h^{1,0}(Y_i) & = -e^{1,0}(Z_{\Theta^{[2]}}\times \left(\C^* \amalg 2pts\right))  \\
 & = \ell^*(\Theta^{[2]})\cdot (e^{0,0}(\C^*)+2e^{0,0}(pt)) \\
 & = \ell^*(\Theta^{[2]}) \, .
 \end{aligned}
\end{equation}
We have $h^{2,0}(Y_i) = e^{2,0}(Y_i) = 0$ as no stratum contributes. Already for the highest stratum
$Z_{\Theta^{[2]}}$, we have to count interior points to 3-dimensional faces of $\Theta^{[2]}$, of which there are none.

\subsubsection{Divisors Interior to 2D faces of \texorpdfstring{$\Delta^\circ$}{Lg}}

Divisors originating from points $\nu_i$ interior to 2-dimensional faces $\Theta^{\circ[2]}$ of $\Delta^\circ$ dual to
1-dimensional faces $\Theta^{[1]}$ of $\Delta$ have a stratification
\begin{equation}
Y_i = Z_{\Theta^{[1]}}\times\left((\C^*)^2 + \sum \C^* + \sum pt \right)  \, ,
\end{equation}
where the $(\C^*)^2$, $\C^*$ and $pt$ originate from $0$, $1$ and $2$ simplices interior to $\Theta^{[2]}$
containing $\nu_i$. As $Z_{\Theta^{[1]}}$ is a 0-dimensional stratum made up of $\ell^*(\Theta^{[1]})+1$ points,
such divisors are reducible with $\ell^*(\Theta^{[1]})+1$ components, each of which is toric (the corresponding variety is
determined by the fan $star(\nu_i))$. Hence $h^{1,0}(Y_i)=h^{2,0}(Y_i)=0$.

\section{Degenerating an Elliptic K3 Surface into a Pair of Rational Elliptic Surfaces}\label{sect:k3stabledeg}

The stable degeneration limit which leads to the decomposition of a K3 surface 
into two rational elliptic surfaces is well-known in string theory due to its relevance to the duality between 
heterotic $E_8\times E_8$ string theory and F-theory \cite{Morrison:1996pp}. Here, we review it for pedagogical reasons as it 
inspires the construction presented in Section \ref{sect:adhocconstruction}.

The simplest presentation starts from a family of elliptic K3 surfaces which are 
hypersurfaces in a toric variety with weight system
\be
\begin{tabular}{ccccc}
$y$ & $x$ & $z$ & $\xi_1$ & $\xi_2$ \\
\hline 
$3$ & $2$ & $1$ & $0$ & $0$ \\
$6$ & $4$ & $0$ & $1$ & $1$
\end{tabular}
\ee
given by an equation
\begin{equation}\label{eq:k3e8e8}
 y^2 = x^3 + \alpha x z^4 \xi_1^4 \xi_2^4  + z^6(\lambda_1 \xi_1^5 \xi_2^7 +  \beta \xi_1^6 
\xi_2^6 + \xi_1^7 \xi_2^5) \, 
\end{equation}
for two complex parameters $\alpha$ and $\beta$. A generic member of this family of K3 surfaces has two singularities 
of type $E_8$ ($II^*$ fibres) over $\xi_1=0$ and $\xi_2 = 0$. 

The stable degeneration into two rational elliptic surfaces is found over $\lambda = 1$ after
performing a weighted blowup (at $y=x=\xi_1=\lambda_1=0)$ with weights 
$(3,2,1,1;-1)$ of this family. The weight system becomes
\be
\begin{tabular}{ccccccc}
$y$ & $x$ & $z$ & $\xi_1$ & $\xi_2$ & $\lambda_1$ & $\lambda_2$ \\
\hline 
$3$ & $2$ & $1$ & $0$ & $0$ & $0$ & $0$\\
$6$ & $4$ & $0$ & $1$ & $1$ & $0$ & $0$ \\
$3$ & $2$ & $0$ & $1$ & $0$ & $1$ & $-1$ 
\end{tabular}
\ee
and the proper transform of \eqref{eq:k3e8e8} is
\begin{equation}\label{eq:k3e8e8bu}
 y^2 = x^3 + \alpha x z^4 \xi_1^4 \xi_2^4  + z^6(\lambda_1 \xi_1^5 \xi_2^7 +  \beta \xi_1^6 
\xi_2^6 + \lambda_2 \xi_1^7 \xi_2^5) \, .
\end{equation}

The two $dP_9$ are now obtained as $\lambda_1=0$ and $\lambda_2=0$. Let us make 
this explicit for $\lambda_2=0$: setting $\lambda_2=0$
allows us to set $\xi_2 =1$, as the two coordinates are in the SR ideal. We are 
hence left with a homogeneous equation of degrees
$(6,6)$, namely 
\begin{equation}\label{eq:dp9}
 y^2 = x^3 + \alpha  x z^4 \xi_1^4   + z^6 \xi_1^5 (\lambda_1 +  \beta \xi_1 ) \, .
 \end{equation}
in an ambient toric space with weights
\be\label{eq:dp9weights}
\begin{tabular}{ccccc}
$y$ & $x$ & $z$ & $\xi_1$ & $\lambda_1$ \\
\hline 
$3$ & $2$ & $1$ & $0$ & $0$   \\
$3$ & $2$ & $0$ & $1$  & $1$ 
\end{tabular}
\ee
which is nothing but a rational elliptic surface $dP_9$. Note that the $E_8$ singularity is preserved in 
this process.

The above situation can of course also be described in terms of tops. Here, we 
would start from two $E_8$ tops over the reflexive polytope associated with $\P_{123}$ \cite{Candelas:1996su} and 
would naturally obtain a resolved version of a K3 surface with two $II^*$ fibres. Using the construction presented in 
Section \ref{sect:adhocconstruction} directly results in \eqref{eq:dp9}.

Even though we have started with an elliptic K3 with two singularities of type 
$E_8$, the same construction also works if we start to unfold the two $E_8$ singularities. E.g. we can deform the $E_8$ singularity 
at $\xi_1=0$ by simply adding terms of lower powers in $\xi_1$ to \eqref{eq:k3e8e8}. This can be done systematically 
(i.e. we automatically find which polynomials we have to add) by considering tops which are smaller than the 
$E_8$ top. Going through the same steps performed above, this will propagate to \eqref{eq:dp9}, i.e. we can specify a $dP_9$ with 
a specific degenerate fibre using the corresponding top. Note that adding monomials which (partially) deform the $E_8$ singularity 
at $\xi_2 = 0$ do not change at all what happens for the $dP_9$ at $\lambda_2 = 0$ as all such terms come with positive powers of $\lambda_2$. 
Again, the resulting families of $dP_9$ surfaces can be directly obtained from the tops by using the construction of Section 
\ref{sect:adhocconstruction}.

\section{Polytopes and Torsion in \texorpdfstring{$H^3$}{Lg}}\label{sect:polybrauer}

In this section we record a list of reflexive polytopes for which the corresponding Calabi-Yau hypersurface
has non-vanishing $\mbox{Tors}( H^3(X,\mathbb{Z}) )$ and give some data on reflexive subpolytopes of 
codimension one, relevant to the discussion in Section \ref{sect:sum_proof}.

Below, we have listed the 16 ${\bf N}$-lattice polytopes $\Delta^\circ$, taken from \cite{KrBat,He:2013ofa}, for which the corresponding reflexive pair $(\Delta,\Delta^\circ)$ gives rise to a Calabi-Yau hypersurface $X_{(\Delta,\Delta^\circ)}$ such that $\pi_1(X_{(\Delta,\Delta^\circ)}) \neq 0$. According to \cite{KrBat}, and consistent with mirror symmetry, the mirror Calabi-Yau manifolds $X_{(\Delta^\circ,\Delta)}$ have non-vanishing
torsion, $\mbox{Tors}( H^3(X_{(\Delta^\circ,\Delta)},\Z) )\neq 0$.  Hence 
\begin{equation}\label{eq:toricbrauerapp}
 \mbox{Hom}(\,\Lambda^2 {\bf N}/({\bf N} \wedge {\bf N}_{\Delta^\circ}^{(2)}), \mathbb{Q}/\mathbb{Z}\,) 
\end{equation}
is non-vanishing only for these 16 cases. For a pair of projecting tops $\Diamond^\circ,\Diamond$ to define a threefold $Z$ with non-vanishing torsion $\mbox{Tors}(H^3(Z,\Z))$,
\begin{equation}\label{torsh3dia}
 \mbox{Hom}(\,\Lambda^2 {\bf N}/({\bf N} \wedge {\bf N}_{\Diamond}^{(2)}), \mathbb{Q}/\mathbb{Z}\,) 
\end{equation}
needs to be non-zero. As we can complete $\Diamond$ together with a copy of itself (reflected on $F$) to a reflexive polytope for which \eqref{torsh3dia}
will be the same, it follows that only tops contained in the reflexive polyhedra $\Delta^\circ$ recorded below can lead to $\mbox{Tors}(H^3(Z,\Z))$.
None of these contains any projecting top, so that we can conclude that $\mbox{Tors}( H^3(Z,\Z) )= 0$
for any $Z$ satisfying the definition of Section \ref{sect:sum_proof}.

\begin{longtable}{cc}

$\left(\begin{array}{rrrrr}
 -5 & 0 & 0 & 0 & 5 \\
 1 & -4 & 0 & 3 & 0 \\
 0 & -2 & 1 & 1 & 0 \\
 -1 & 1 & 0 & -1 & 1
 \end{array}\right)$\, , & $\left(\begin{array}{rrrrrr}
 -3 & 0 & 0 & 0 & 0 & 3 \\
 -1 & -1 & 0 & 0 & 0 & 2 \\
 0 & -1 & -1 & 0 & 1 & 1 \\
 1 & -1 & -1 & 1 & 0 & 0
 \end{array}\right)$\, , \\
 &\\
 $\left(\begin{array}{rrrrrrrr}
 -2 & 0 & 0 & 0 & 0 & 0 & 0 & 2 \\
 0 & -1 & -1 & 0 & 0 & 1 & 1 & 0 \\
 0 & -1 & 0 & -1 & 1 & 0 & 1 & 0 \\
 -1 & -1 & 0 & 0 & 0 & 0 & 1 & 1
 \end{array}\right)$\, ,& $\left(\begin{array}{rrrrr}
 -3 & 3 & 0 & 0 & 0 \\
 -1 & -2 & 0 & 0 & 1 \\
 -2 & -1 & 0 & 1 & 0 \\
 1 & -2 & 1 & 0 & 0
 \end{array}\right)$\, , \\
 & \\
 
  $\left(\begin{array}{rrrrr}
 -4 & 4 & 0 & 0 & 0 \\
 -1 & 2 & -1 & 0 & 0 \\
 0 & 1 & -2 & 0 & 1 \\
 -3 & 0 & -1 & 1 & 0
 \end{array}\right)$\, , &$\left(\begin{array}{rrrrrr}
 -4 & 2 & 0 & 0 & 0 & 0 \\
 -3 & 0 & -2 & -1 & 0 & 1 \\
 1 & 0 & -1 & -1 & 1 & 0 \\
 -1 & 1 & 0 & -1 & 0 & 0
 \end{array}\right)$\, , \\
 & \\
 
 $\left(\begin{array}{rrrrr}
 -4 & 2 & 0 & 0 & 0 \\
 -3 & 0 & 1 & -1 & 0 \\
 -7 & 0 & 0 & -1 & 1 \\
 -1 & 1 & 0 & -1 & 0
 \end{array}\right)$\, ,& $\left(\begin{array}{rrrrrr}
 -2 & 0 & 0 & 0 & 4 & 4 \\
 -2 & -1 & 0 & 1 & 1 & 0 \\
 -1 & -1 & 0 & 0 & 3 & 2 \\
 -1 & 0 & 1 & 0 & 2 & 1
 \end{array}\right)$\, , \\
 &\\
$ \left(\begin{array}{rrrrrr}
 -4 & 4 & 0 & 0 & 0 & 0 \\
 -1 & 2 & -1 & 0 & 0 & 0 \\
 0 & 1 & -2 & 0 & 0 & 1 \\
 1 & 0 & -1 & -1 & 1 & 0
 \end{array}\right)$\, ,& $\left(\begin{array}{rrrrrr}
 -4 & 2 & 0 & 0 & 0 & 0 \\
 1 & 0 & -1 & -1 & 0 & 2 \\
 -1 & 1 & -1 & 0 & 0 & 0 \\
 0 & 0 & 0 & -1 & 1 & 1
 \end{array}\right)$ \, ,\\
&\\
$ \left(\begin{array}{rrrrrrr}
 -2 & 0 & 2 & 0 & 0 & 0 & 0 \\
 0 & -1 & 0 & -1 & 0 & 0 & 1 \\
 0 & -1 & 0 & 1 & -1 & 1 & 0 \\
 -1 & -1 & 1 & 1 & 0 & 0 & 0
 \end{array}\right)$\, ,& $\left(\begin{array}{rrrrrr}
 -2 & 0 & 0 & 0 & 2 & 0 \\
 0 & -3 & 1 & -1 & 0 & 0 \\
 0 & -1 & 0 & -1 & 0 & 1 \\
 -1 & 1 & 0 & -1 & 1 & 0
 \end{array}\right)$\, ,\\
 &\\
$ \left(\begin{array}{rrrrrrr}
 -2 & 0 & 0 & 0 & 2 & 0 & 0 \\
 0 & -1 & -1 & 0 & 0 & 2 & 3 \\
 -1 & -1 & 0 & 0 & 1 & 0 & 1 \\
 0 & 0 & -1 & 1 & 0 & 1 & 2
 \end{array}\right)$\, ,& $\left(\begin{array}{rrrrrrr}
 -2 & 0 & 0 & 0 & 0 & 0 & 2 \\
 2 & -1 & -1 & 0 & 2 & 3 & 0 \\
 -1 & -1 & 0 & 0 & 0 & 1 & 1 \\
 2 & 0 & -1 & 1 & 1 & 2 & 0
 \end{array}\right)$\, ,\\
 &\\
$ \left(\begin{array}{rrrrr}
 -4 & 4 & -4 & 0 & 0 \\
 -3 & 2 & -1 & 0 & 0 \\
 -2 & 1 & -2 & 1 & 0 \\
 -1 & 0 & -1 & 0 & 1
 \end{array}\right)$\, ,& $\left(\begin{array}{rrrrrr}
 -4 & -4 & 4 & 0 & 0 & 0 \\
 -3 & -2 & 2 & 0 & 0 & 1 \\
 -2 & -2 & 1 & 0 & 1 & 0 \\
 -1 & -1 & 0 & 1 & 0 & 1
 \end{array}\right)$\, .\\

\end{longtable}

\end{document}